%% file: neurips_2020.tex
\documentclass{article}

\usepackage[final, nonatbib]{neurips_2020}

\usepackage[utf8]{inputenc} % allow utf-8 input
\usepackage[T1]{fontenc}    % use 8-bit T1 fonts
\usepackage{hyperref}       % hyperlinks
\usepackage{url}            % simple URL typesetting
\usepackage{booktabs}       % professional-quality tables
\usepackage{amsfonts}       % blackboard math symbols
\usepackage{nicefrac}       % compact symbols for 1/2, etc.
\usepackage{microtype}      % microtypography
\usepackage{lipsum}
\usepackage{bm}
\usepackage{amsmath}
\usepackage{algorithm}
\usepackage[noend]{algpseudocode}
\usepackage{graphicx}
\usepackage{dsfont}
\usepackage[mathscr]{eucal}
\usepackage{xcolor}
\usepackage{appendix}
\usepackage{mathrsfs} 
\usepackage{enumitem}
\usepackage{nicefrac}
\usepackage{amstext}
\usepackage{verbatim}
\usepackage{xr}
\hypersetup{
	colorlinks = true,
	linkcolor = blue,
	anchorcolor = blue,
	citecolor = magenta,
	filecolor = blue,
	urlcolor = blue
}
\usepackage[normalem]{ulem}

\definecolor{ao}{rgb}{0.0, 0.5, 0.0}

\newtheorem{assumption}{Assumption}

\newtheorem{prop}{Proposition}
\newtheorem{property}{Property}
\newtheorem{definition}{Definition}

\newcounter{protocol}


\newcommand\underrel[2]{\mathrel{\mathop{#2}\limits_{#1}}}

\title{Community detection in sparse time-evolving graphs with a dynamical Bethe-Hessian
}

\author{%
	Lorenzo Dall'Amico \\
	GIPSA-lab, UGA, CNRS, Grenoble INP\\
	\texttt{lorenzo.dall-amico@gipsa-lab.fr} \\
	 \AND
	 Romain Couillet \\
	 GIPSA-lab, UGA, CNRS, Grenoble INP\\
	 L2S, CentraleSupélec, University of Paris Saclay
	 \And
	 Nicolas Tremblay \\
	 GIPSA-lab, UGA, CNRS, Grenoble INP
}

\begin{document}
	
	\maketitle
	
	\begin{abstract}
	This article considers the problem of community detection in sparse dynamical graphs in which the community structure evolves over time. A fast spectral algorithm based on an extension of the Bethe-Hessian matrix is proposed, which benefits from the positive correlation in the class labels and in their temporal evolution and is designed to be applicable to any dynamical graph with a community structure. Under the dynamical degree-corrected stochastic block model, in the case of two classes of equal size, we demonstrate and support with extensive simulations that our proposed algorithm is capable of making non-trivial community reconstruction as soon as theoretically possible, thereby reaching the optimal \emph{detectability threshold} and provably outperforming competing spectral methods.
	\end{abstract}
	
	\section{Introduction}
	\label{sec:introduction}
	
	Complex networks are a powerful tool to describe pairwise interactions among the members of a multi-agent system \cite{barabasi2016network}. One of the most elementary tasks to be performed on networks is community detection \cite{girvan2002community,fortunato2010community}, \emph{i.e.}, the identification of a non overlapping partition of the members (or nodes) of the network, representing its ``mesoscale'' structure. 
    Although most of the attention is still focused on 	\emph{static} community detection \cite{fortunato2010community}, many real networks are intimately dynamic: this is the case of networks representing physical proximity of mobile agents, collaboration interactions in the long run, biological and chemical evolution of group members, etc.\@ (see \cite{holme2015modern} for a review).

	There are many ways to define the concept of communities, particularly in dynamical networks (see \emph{e.g.}, \cite{fortunato2010community, rossetti2018community}). In this article, we focus on 
	the \emph{dynamical degree corrected stochastic block model} (D-DCSBM), formally defined in Section~\ref{sec:model}, which is a variation of the static DCSBM \cite{karrer2011stochastic,abbe2017community}. Specifically, letting $\mathcal G_t$ be the $k$-community graph at time instant $t$, $\mathcal G_t$ is generated independently of $\mathcal G_{t'}$ for all $t'\neq t$, but only a fraction $1-\eta$ (for $\eta\in[0,1]$) of the nodes changes class association between time $t$ and time $t+1$. The degree correction lets the nodes have an arbitrary degree distribution, thereby possibly accounting for the broad distributions typical of real networks \cite{barabasi1999emergence}. Of fundamental importance, in the static regime, two-class DCSBM graphs exhibit a \emph{detectability threshold} below which no algorithm can asymptotically find a node partition non trivially aligned to the genuine classes \cite{decelle2011asymptotic,massoulie2014community,mossel2012stochastic,gulikers2016non,gulikers2018impossibility}. Under a D-DCSBM model, one can similarly define a \emph{dynamical detectability threshold} which considers the inference problem on the graph sequence $\{\mathcal{G}_t\}_{t=1,\dots,T}$ \cite{ghasemian2016detectability}. For $k>2$ classes, the identification of a detectability threshold remains an open problem. 
	
	\medskip
	
	Spectral clustering is arguably one of the most successful ways to perform community detection \cite{von2007tutorial}. Instances of spectral methods are indeed known to attain the detectability threshold in various contexts (in dense \cite{couillet2016kernel,ali2018random} or sparse \cite{gulikers2018impossibility,krzakala2013spectral,saade2014spectral,bordenave2015non} stochastic block models) and are experimentally observed to perform competitively with the Bayes optimal solution \cite{krzakala2013spectral,saade2014spectral}. Recently, spectral clustering algorithms have also been explored in the dynamic regime \cite{chi2007evolutionary,qin2016multi,chi2009evolutionary,liu2018global,pensky2019spectral, keriven2020sparse}.

	Two of the major pitfalls of dynamical spectral methods are \emph{sparsity}, when the node degrees do not scale with the size $n$ of the graph, and \emph{small label persistence}, when the fraction of nodes $1-\eta$ that change label at any time instant $t$ is of order $O_n(1)$.
	Small persistence realistically assumes that, successive observations of the graph being independent across time, their community configuration must also evolve non-trivially.
	Under a sparse regime, but for $1-\eta = o_n(1)$, \cite{keriven2020sparse} suggests to average the adjacency matrices over multiple time instances to obtain efficient community reconstruction. To the best of our knowledge, the work of \cite{ghasemian2016detectability} provides the only existing spectral algorithm properly treating both sparsity and small label persistence. In the spirit of \cite{krzakala2013spectral}, the proposed method arises from a linearization of the (asymptotically optimal) \emph{belief propagation} algorithm (BP), which is capable of obtaining non-trivial partitions (\emph{i.e.}, better than random guess) as soon as theoretically possible. However, their resulting dynamical non-backtracking matrix depends on an \emph{a priori} unknown parameter\footnote{In order to design their dynamical non-backtracking matrix, the average number of connections among nodes in the same and across communities must be known.}, so the algorithm is practically inapplicable. 
	
	\medskip
	
	As an answer to these limitations, this article proposes a new spectral algorithm adapted to the sparse regime, which is able to detect communities even under little (or no) persistence in the community labels and which benefits from persistence to improve classification performance over a static algorithm run independently at each time-step. Specifically,
	\begin{enumerate}[leftmargin=0.15in]
		\item We introduce a dynamical Bethe-Hessian matrix which, for $k=2$, retrieves non-trivial communities as soon as theoretically possible. 
		As a by-product, we offer new results on the spectrum of the dynamical non-backtracking of \cite{ghasemian2016detectability}.
        \item We provide an algorithm applicable to any graph with $k\geq2$ communities of arbitrary sizes.\footnote{The algorithm \emph{a priori} requires that $\eta$ be known; otherwise, $\eta$ can be estimated through cross-validation.}. On top of Python codes to reproduce most of the figures of this paper (available in the supplementary material), we provide an efficient Julia implementation, part of the CoDeBetHe package (community detection with the Bethe-Hessian), available at \href{https://github.com/lorenzodallamico}{github.com/lorenzodallamico}.
  	\end{enumerate}

	\textbf{Notations.}
	Function $\mathds{1}_{x}$ is the indicator equal to $1$ if condition $x$ is verified and $0$ otherwise.
	Column vectors are indicated in bold ($\bm{v}$), matrices ($M$) and vector elements ($v_i$) in standard font. Vector $\bm{1}_n\in\mathbb R^n$ is the all-ones vector. The index $t$ always refers to time. The set $\partial i = \{j : (i,j)\in \mathcal{E}\}$ are the neighbors of $i$ in graph $\mathcal G=(\mathcal V,\mathcal E)$ with edge set $\mathcal E$. The spectral radius of matrix $M$ is $\rho(M)$.

\section{Model and setting}
	\label{sec:model}
	
	Let $\mathcal \{\mathcal{G}_t\}_{t = 1,\dots,T}$ be a sequence of unweighted and undirected graphs, each with $n$ nodes. At time step $t$, $\mathcal{E}_t$ and $\mathcal{V}_t$ denote the set of edges and nodes, respectively, which form $\mathcal{G}_t$, with $\mathcal{V}_t \cap \mathcal{V}_{t'} = \emptyset$, for $t'\neq t$: each node has $T$ copies, each copy being a different object. We denote with $i_t$, for $1 \leq i \leq n$ and $1 \leq t \leq T$, a node in $\mathcal{V}_t$.  We call $A^{(t)} \in \{0,1\}^{n \times n}$ the symmetric adjacency matrix of $\mathcal{G}_t$, defined as $A_{ij}^{(t)} = \mathds{1}_{(ij)\in\mathcal{E}_t}$, and $D^{(t)}={\rm diag}(A^{(t)}\bm{1}_n) \in \mathbb{N}^{n\times n}$ its associated degree matrix. We now detail the generative model for $\{\mathcal{G}_t\}_{t=1,\dots, T}$.
	
	\subsection{The dynamical degree corrected stochastic block model}
	
	For readability, until Section~\ref{sec:algo}, where among other generalizations, we will consider graphs with an arbitrary number of classes $k$, we focus on a model with two classes of equal size. Let $\ell_{i_t} \in \{1,2\}$ be the label of node $i_t$. The vector $\{\ell_{i_{t=1}}\}_{i=1,\dots,n}$ is initialized by assigning random labels ($1$ or $2$) with equal probability. The labels are then updated for $2 \leq t \leq T$ according to the Markov process
	\begin{align}
	\ell_{i_{t}} =   \begin{cases}
	\ell_{i_{t-1}}  &\text{w.p.\@ } \eta \\
	a &\text{w.p.\@ } \frac{1-\eta}{2}, ~~a \in \{1,2\},
	\end{cases}
	\label{eq:label_markov}
	\end{align}
	\emph{i.e.}, the label of node $i_t$ is maintained with probability $\eta$ and otherwise reassigned at random with probability $1-\eta$.
	Note that a proportion of the reassigned nodes from time $t$ will be affected the same labels at time $t+1$.
	The entries of the adjacency matrix $A^{(t)}$ of $\mathcal{G}_t$ are generated independently and independently across $t$, according to:
	\begin{align}
	\label{eq:A}
	\mathbb{P}(A^{(t)}_{ij} = 1) = \theta_{i}\theta_{j}\frac{C_{\ell_{i_t},\ell_{j_t}}}{n}, \quad \forall~i>j.
	\end{align}
	The vector $\bm{\theta}=(\theta_1,\ldots,\theta_n)$ enables to induce any arbitrary degree distribution and satisfies $\frac{1}{n}\sum_{i =1}^n \theta_{i} = 1$ and $\frac{1}{n}\sum_{i=1}^n \theta_{i}^2 \equiv \Phi = O_n(1)$. The matrix $C \in \mathbb{R}^{2\times 2}$ contains the class affinities with $C_{a=b} \equiv c_{\rm in}$ and $C_{a\neq b} \equiv c_{\rm out}$, $c_{\rm in}$ and $c_{\rm out}$ being independent of $n$. The expected average graph degree is $c \equiv (c_{\rm in} + c_{\rm out})/2 = O_n(1)$ assumed to satisfy $c\Phi>1$: according to \eqref{eq:label_markov}--\eqref{eq:A}, this is the necessary (and sufficient) condition such that, at each time step, $\mathcal{G}_t$ has a giant component\footnote{The existence of a giant component at each time $t$ ensures a well-defined community detection problem when $n\to\infty$. In practice, $\mathcal{G}_t$ will typically be the union of a giant connected sub-graph, in which all communities are represented, and a few isolated nodes. These isolated nodes can be understood as nodes of a network absent at time $t$. In this sense, the D-DCSBM is suitable to model dynamic networks with varying size across time.} \cite{dall2020unified}. This condition imposes constraint on $c$, hence on how sparse the graphs $\{\mathcal{G}_{t}\}_{t=1,\dots,T}$ can be.
	
	\medskip
	
	We insist that the process \eqref{eq:label_markov}--\eqref{eq:A} builds on a dual time-scale assumption: a short range governing the evolution of graph edges (reconfigured at each time step) and a long range governing the evolution of communities. The article mainly focuses on the long range evolution as independent realizations of $\mathcal{G}_t$ are assumed at successive times. Appendix~\ref{app:memory} discusses the extension of this framework to $\mathcal{G}_t$ evolving slowly with time, thereby allowing for \emph{edge persistence} across time.
	
	\medskip
		
\begin{figure}[t!]
		\centering
		\includegraphics[width = 0.45\columnwidth]{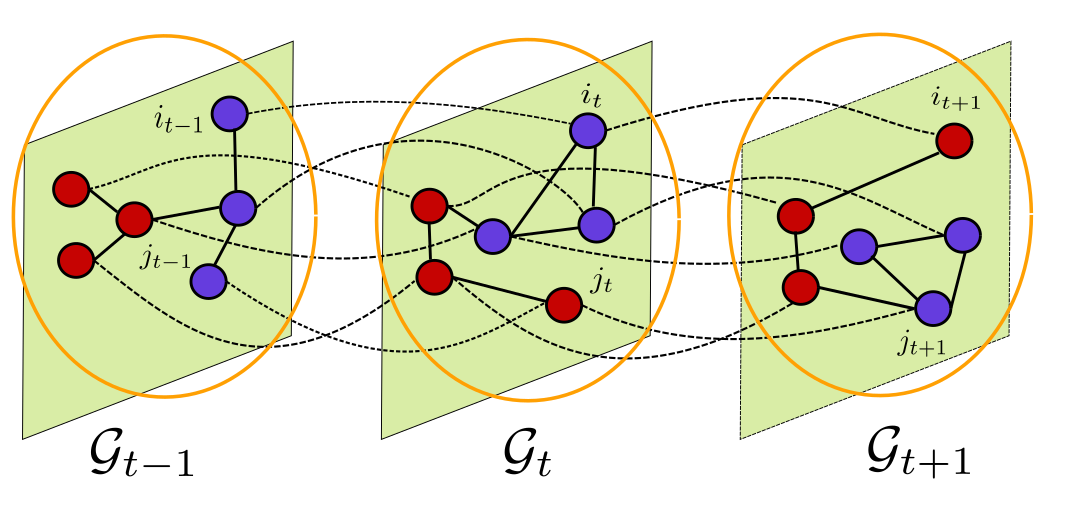}
		\caption{Three successive instances of a dynamical network $\mathcal{G}$. Classes are emphasized by node colors and can evolve with time. Network edges, that change over time, are indicated in solid lines, while ``temporal edges'' in dashed lines connect each graph to its temporal neighbors. Nodes of a common time step are circled in orange.}
		\label{fig:G}
	\end{figure}

	Our objective is to solve the problem of community reconstruction on the \emph{dynamical graph} $\mathcal{G}$ constructed, as illustrated in Figure~\ref{fig:G}, from the $T$ independent instances $\{\mathcal{G}_t\}_{t = 1,\dots,T}$.
	
	\begin{definition}
		Letting $\{\mathcal{G}_t\}_{t = 1\dots T}$ be a sequence of graphs independently generated from \eqref{eq:label_markov}--\eqref{eq:A}, $\mathcal G=\mathcal{G}(\mathcal{V},\mathcal{E})$ is the graph with $\mathcal{V} = \cup_{t = 1}^{T} \mathcal{V}_t$ and $\mathcal{E} = \left(\cup_{t = 1}^T \mathcal{E}_t\right) \cup\left(\cup_{t = 1}^{T-1}\cup_{i = 1}^n (i_t,i_{t+1})\right)$. The adjacency and degree matrices of $\mathcal{G}$ are denoted with $A, D \in \mathbb{N}^{nT\times nT}$, respectively. In other words, the graphs $\mathcal{G}_t$ are joined adding extra edges between the nodes $i_t$ and their temporal neighbors $i_{t\pm1}$.
		\label{def:G}
	\end{definition}

	\subsection{Detectability threshold in the D-DCSBM}
	\label{subsec:model.detectability}
	
    Let $\lambda=(c_{\rm in}-c_{\rm out})/(c_{\rm in}+c_{\rm out})$ be the co-variance between neighboring labels \cite{gulikers2018impossibility,dall2019revisiting}. Based on \cite{janson2004robust}, the authors of \cite{ghasemian2016detectability} conjecture that, for the D-SBM (for which $\theta_{i} = 1$ for all $i$), as $n, T \to \infty$, non-trivial class reconstruction is feasible if and only if $\alpha \equiv \sqrt{c\lambda^2} >\alpha_c(\infty,\eta)$, where $\alpha_c(\infty,\eta)$ is the \emph{detectability threshold} defined as the unique value of $\bar{\alpha}>0$ for which the largest eigenvalue of
    	\begin{align}
    	M_{\infty}(\bar{\alpha},\eta) = \begin{pmatrix}
    	\bar{\alpha}^2 & 2\eta^2 \\
    	\bar{\alpha}^2 & \eta^2
    	\end{pmatrix}
    	\label{eq:matrix_M}
    	\end{align}
    is equal to one. Inspired by~\cite{gulikers2018impossibility} who adapted the detectability condition to the DCSBM model in the static case, we show (see Appendix~\ref{app:detectability}) that this result can be extended to the D-DCSBM by (i)~redefining $\alpha$ as $\alpha \equiv \sqrt{c\Phi\lambda^2}$ and (ii)~for finite $T$ (but $n \to \infty$) by redefining $\alpha_c(T,\eta)$ as the value of $\bar{\alpha}$ for which the largest eigenvalue of
    	\begin{align}
    		M_T(\bar{\alpha},\eta) = \begin{pmatrix}
    		M_d & M_+ & 0  & \dots&0 \\
    		M_- & M_d & \ddots &\dots &0 \\
    		0&M_{-}&\ddots&M_+&0&\\
    		\vdots&\vdots&\ddots&M_d&M_+\\
    		0&0&\dots& M_-&M_d
    		\end{pmatrix}, 
    		\text{  where  }
    		\begin{cases}
    		&M_d = \left(\begin{smallmatrix}
    		0&0&0 \\
    		\eta^2 & \bar{\alpha}^2 & \eta^2 \\
    		0&0&0
    		\end{smallmatrix}\right),\\[10pt]
    		&M_+ = \left(\begin{smallmatrix}
    		0&0&0 \\
    		0&0&0 \\
    		0 & \bar{\alpha}^2 & \eta^2
    		\end{smallmatrix}\right),\\[10pt]
    		&M_- = \left(\begin{smallmatrix}
    		\eta^2&\bar{\alpha}^2&0 \\
    		0&0&0 \\
    		0&0&0
    		\end{smallmatrix}\right),
    		\end{cases}
    		\label{eq:transition}
    		\end{align}
    is equal to one. The detailed derivation of $M_T(\bar{\alpha},\eta)$ are reported in Appendix~\ref{app:detectability}, which provides an explicit expression to $\alpha_c(T,\eta)$, following the arguments of \cite{ghasemian2016detectability}. The definition of $M_T(\bar{\alpha},\eta)$ is more elaborate than $M_{\infty}(\bar{\alpha},\eta)$ due to the finite-time structure of $\mathcal{G}$: each node $i_t$ has two temporal connections with $i_{t+1}$ and $i_{t-1}$, except for the ``time boundary'' nodes of $\mathcal{G}_{t=1}$ and $\mathcal{G}_{t=T}$. As $T\to \infty$, these boundaries can be neglected and the leading eigenvalue of $M_T( \bar{\alpha},\eta)$ reduces to that of $M_\infty( \bar{\alpha},\eta)$.
    The expression of $\alpha_c(T,\eta)$ can be computed analytically for $T=2,3,4$ and $T\to \infty$:
    	\begin{align}
    		\alpha_c(T=2,\eta) &= \left(1+\eta^2\right)^{-\frac{1}{2}}; \qquad \alpha_c(T=3,\eta) = \sqrt{2}\left(2+\eta^4+\eta^2\sqrt{8+\eta^4}\right)^{-\frac{1}{2}}\\ \alpha_c(T = 4, \eta) &= \sqrt{2}\left(2+\eta^2+\eta^6+\eta\sqrt{\eta^8+2\eta^4+8\eta^2+5}\right)^{-\frac{1}{2}};~~ \alpha_c(\infty, \eta) = \left(\frac{1+\eta^2}{1-\eta^2} \right)^{-\frac{1}{2}}.\nonumber
    		\end{align}
    For other values of $T$, $\alpha_c(T,\eta)$ is best evaluated numerically. For all $T$: (i)~if $\eta = 0$ (no correlation among the labels), one recovers $\alpha_c = 1$, the transition's position in the static DCSBM~\cite{gulikers2018impossibility}, as expected; (ii)~if $\eta = 1$, $\alpha_c = 1/\sqrt{T}$, the static threshold obtained by averaging the adjacency matrix over its $T$  independent and identically distributed realizations. We also numerically confirm that for all $T$, $\alpha_c(T,\eta)$ is a decreasing function of $\eta$: higher label persistence allows to solve harder problems.

    	\section{Main results}
    	\label{sec:main}
    	This section develops a new ``dynamical'' Bethe-Hessian matrix associated to the graph $\mathcal{G}$, for which we show there exists \emph{at least} one eigenvector (recall that $k=2$ classes so far) strongly aligned to the community labels if $\alpha > \alpha_c(T,\eta)$, thereby allowing for high performance community detection down to the detectability threshold. % $\alpha=\alpha_c(T,\eta)$. 
    	The eigenvectors containing information can be up to $T$, but only one of them is guaranteed to exist when $\alpha > \alpha_c(T,\eta)$ and it can alone reconstruct communities.
    	
    	\begin{figure}[t!]
    		\centering
    		\includegraphics[width=\columnwidth]{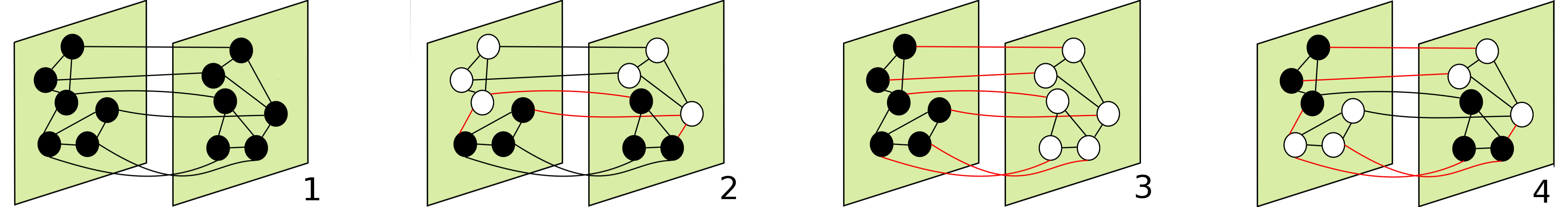}
    		\caption{Sketch of the $4$ stable modes for two communities and $T=2$. In black we indicate the direction $s_i > 0$, in white $s_i < 0$. The red edges correspond to the frustrated edges connecting spins with opposite direction.}
    		\label{fig:stable}
    	\end{figure}
    	
	\subsection{The dynamical Bethe-Hessian matrix}
	\label{subsec:main.BH}

	As in \cite{saade2014spectral,dall2019revisiting}, our approach exploits a statistical physics analogy between the modelling of spontaneous magnetization of spins with \emph{ferromagnetic} interaction \cite{mezard2009information} and the modelling of communities of nodes in sparse graphs. We attach to each node a \emph{spin} variable $s_{i_t} \in \{\pm 1\}$, for $1\leq i\leq n$ and $1\leq t\leq T$. The energy of a spin configuration $\bm{s}\in\{\pm 1\}^{nT}$ is given by the \emph{Hamiltonian}
	\begin{align}
	\mathcal{H}_{\xi,h}(\bm{s}) = -\sum_{t = 1}^T\left( \sum_{(i_t,j_t)\in\mathcal{E}_t}{\rm ath}(\xi)~ s_{i_t}s_{j_t} + \sum_{i_t \in \mathcal{V}_t}{\rm ath}(h)~ s_{i_t}s_{i_{t+1}}\right)
	\label{eq:ising}
	\end{align}
	with $s_{i_{T+1}} = 0$ by convention. Here, the coupling constants $\xi, h \in [0,1)$ modulate the interaction among nodes at time $t$ and between the same node at time instants $t$ and $t+1$, respectively, and appear inside inverse hyperbolic tangents for notational ease. Intuitively, the spin vector $\bm{s}$ can be mapped to the class affiliation vector $\bm{\sigma} = 2\bm{\ell}-3$. The first term in the main parenthesis of \eqref{eq:ising} favors configurations in which neighboring nodes have the same label, while the second term favors configurations in which the label is kept across successive time instants. This last term enforces persistence in the community evolution. 
	
	\medskip
	
	The configurations $\bm{s}$ representing the local minima of $H_{\xi,h}(\bm{s})$ are determined by the mesoscale structure of $\mathcal{G}$ and are sketched for $T=2$ in Figure~\ref{fig:stable}. The lowest energy state corresponds to $\bm{s} = \bm{1}_{nT}$: this is the non-informative \emph{ferromagnetic configuration}. Similarly, mode 3 of Figure~\ref{fig:stable} groups together nodes in the same community and is equally useless for reconstruction. On the opposite, modes 2 and 4 of Figure~\ref{fig:stable} divide the nodes according to the class structure of $\mathcal{G}$ and can be used for community reconstruction. In general, for $k$ classes and $T>2$ time frames, $kT$ local minima arise, mixing together time and class clusters. Note importantly that mode 1 always has a lower energy than mode 3 and mode 2 a lower energy than mode 4. However, the ordering of energies of modes 2 and 3 is in general not \emph{a priori} known. We will further comment on this remark which has important consequences for the subsequent analysis as well as for the design of our proposed community detection algorithm. 	

	\medskip
	
	We show in Appendix~\ref{app:BH} that these lowest energy modes can be approximated by the eigenvectors associated with the smallest eigenvalues of the \emph{Bethe-Hessian} matrix $H_{\xi,h}\in\mathbb{R}^{nT\times nT}$, defined by
	\begin{align}
	\label{eq:BHMatrix}
	\left(H_{\xi,h}\right)_{i_t,j_{t'}} = \begin{cases}
	\left(\frac{\xi^2D^{(t)}-\xi A^{(t)}}{1-\xi^2} + \frac{1+h^2(\phi_t-1)}{1-h^2}I_n\right)_{ij} \quad &{\rm if~} t = t' \\
	\left(-\frac{h}{1-h^2}I_n\right)_{ij} &{\rm if}~ t = t' \pm 1,
	\end{cases}
	\end{align}
	in which $\phi_t = 1$ if $t = 1$ or $t = T$ and $\phi_t = 2$ otherwise. The aforementioned lack of a precise knowledge of the relative position of the informative modes in the energy spectrum of the Hamiltonian hampers the identification of the position of the corresponding informative eigenvectors of $H_{\xi,h}$. This is of major importance when designing a spectral clustering algorithm based on $H_{\xi,h}$.

	\subsection{Community detectability with the dynamic Bethe-Hessian} 
	\label{subsec:main.th}
	
	We thus now turn to our main result (Proposition~\ref{prop:2}), whose theoretical support is given in Appendix~\ref{app:proof}, centered on the question of appropriately choosing a pair $(\xi,h)$ which ensures non-trivial community detection with $H_{\xi,h}$ as soon as $\alpha > \alpha_c(T,\eta)$ and which, in addition, necessarily exploits the informative eigenvectors of $H_{\xi,h}$ without knowing their precise location in the spectrum.
	
	\medskip
	
	Let us first introduce an important intermediary object: the weighted non-backtracking matrix $B_{\xi,h}$, defined on the set of \emph{directed} edges $\mathcal{E}^d$ of $\mathcal{G}$. Letting $\omega_{ij} = \xi$ if there exists a time instant $t$ such that nodes $i,j$ belong to $\mathcal{V}_t$, and $\omega_{ij} = h$ for time edges, the entries $(ij),(kl) \in \mathcal{E}^d$ of $B_{\xi,h}$ are defined as
	\begin{align}
	\left(B_{\xi,h}\right)_{(ij)(kl)} = \mathds{1}_{jk}(1-\mathds{1}_{il})~ \omega_{kl}.
	\label{eq:matrix_B}
	\end{align}
	The spectra, and notably the isolated eigenvalues and their associated eigenvectors, of the matrices $B_{\xi,h}$ and $H_{\xi,h}$ have important common properties \cite{terras2010zeta, watanabe2009graph}. 
	As $n\to\infty$, both the spectra of the Bethe-Hessian and non-backtracking matrices are the union of isolated eigenvalues (the eigenvectors of which carry the information on the mesoscale structure of $\mathcal{G}$) and of a bulk of uninformative eigenvalues \cite{bordenave2015non, gulikers2016non}. This relation allows us to establish the following key result.

	\begin{prop}
		\label{prop:2}
		Let $\lambda_d = \frac{\alpha_c(T,\eta)}{\sqrt{c\Phi}}$. Then, as $n \to \infty$, (i) the complex eigenvalues forming the bulk spectrum of $B_{\lambda_d,\eta}$ are asymptotically bounded within the unit disk  (ii) the smallest eigenvalues of the (real) bulk spectrum of $H_{\lambda_d,\eta}$ tend to $0^+$
		and (iii) the number of isolated negative eigenvalues of $H_{\lambda_d,\eta}$ is equal to the number of real isolated eigenvalues of $B_{\lambda_d,\eta}$ greater than $1$.
		
		In particular, if $\alpha > \alpha_c(T,\eta)$, at least one of the isolated real eigenvalues of $B_{\lambda_d,\eta}$ larger than $1$ and one of the negative isolated eigenvalues of $H_{\lambda_d,\eta}$ are informative in the sense that their associated eigenvectors are correlated to the vector of community labels.
	\end{prop}

    Proposition~\ref{prop:2} indicates that, if $\alpha > \alpha_c(T,\eta)$, 
    certainly there is one informative eigenvector (more precisely, mode 2 of Figure~\ref{fig:stable}) which is associated with \emph{one of the few isolated negative eigenvalues} of $H_{\lambda_d,\eta}$. Other informative eigenvectors (\emph{e.g.} mode 4 of Figure~\ref{fig:stable}) may be associated to negative eigenvalues of $H_{\lambda_d,\eta}$, but their existence is not guaranteed.
    By performing spectral clustering on these few negative eigenvalues and appropriately handling the size-$nT$ eigenvectors, one can then be assured to extract the desired community information. We empirically confirm that using \emph{all} the eigenvectors associated with the isolated negative eigenvalues (instead of only the desired informative eigenvector with unknown location) to form a \emph{low dimensional vector embedding} of the nodes is redundant but it does not severely compromise the performance of the final k-means step of the standard spectral clustering method \cite{lloyd1982least}. The choice $\xi = \lambda_d$ and $h=\eta$ therefore almost immediately induces an explicit algorithm applicable to arbitrary networks and which, as later discussed in Section~\ref{sec:algo}, straightforwardly extends to graphs with $k>2$ communities.

    To best understand the structure of $H_{\lambda_d,\eta}$, a further comment should be made on the expected number of its negative eigenvalues. It may in particular be shown that, in the limit $\eta \to 0$, the off-diagonal blocks of $H_{\lambda_d,\eta}$ vanish and exactly $2T$ negative eigenvalues get isolated, the $T$ smallest negative being almost equal and uninformative and the latter $T$ almost equal but informative. In the limit $\eta \to 1$ instead, the configurations alike modes 3 and 4 of Figure~\ref{fig:stable} are energetically penalized (recall \eqref{eq:ising}) and do not produce any isolated eigenvalue, thus $H_{\lambda_d,\eta}$ only has two negative eigenvalues.

    Appendix~\ref{app:proof} shows that a better choice for $\xi$ is in fact $\lambda$, instead of $\lambda_d$. Experimental verification confirms that, as in the static regime \cite{dall2019revisiting}, this is due to the fact that, unlike $H_{\lambda,\eta}$, the entries of the informative eigenvectors of $H_{\lambda_d,\eta}$ are tainted by the graph degrees, thereby \emph{distorting to some extent} the class information.\footnote{In the present symmetric $k=2$ setting, one expects the entries of the informative eigenvector to be noisy versions of $\pm 1$ values in which the degree dependence intervenes only in the variance, but not in the mean 
    (see \cite{dall2019revisiting,dall2020unified} for a thorough study in the static case).
    For $\xi=\lambda_d$ though, 
    the mean itself depends on the node degree and impedes the performance of k-means}. On the opposite, the eigenvector of $H_{\lambda,\eta}$ associated to the eigenvalue closest to zero (which in this case is isolated while the bulk is away from zero) is informative but \emph{not tainted} by the graph degree heterogeneity. Although both choices of $\xi$ provably enable non-trivial community recovery down to the threshold, $\xi=\lambda$ is expected to outperform $\xi=\lambda_d$, especially as $\alpha$ increases away from the threshold.
    Consequently, if one has access to prior knowledge on $\lambda$, then the eigenvectors of $H_{\lambda,\eta}$ should be used for best performance. However, in practice, providing a good estimate of $\lambda$ in reasonable time remains a challenge, especially for $k\geq 2$. This is why we prefer the choice $\xi=\lambda_d$, as $\lambda_d$ is an explicit function of $\alpha_c(T,\eta)$, $c$ and $\Phi$ all of which can be easily estimated.
    
\section{Algorithm and performance comparison}
	\label{sec:algo}
	
	These discussions place us in a position to provide an algorithmic answer to the dynamic community detection problem under study. The algorithm, Algorithm~\ref{alg:0}, is shown here to be applicable, up to a few tailored adjustments, to arbitrary real dynamical graphs.

	\subsection{Algorithm implementation on arbitrary networks}
	
	We have previously summarized the main ideas behind a dynamical version of spectral clustering based on $H_{\lambda_d,\eta}$. These form the core of Algorithm~\ref{alg:0}. Yet, in order to devise a practical algorithm, applicable to a broad range of dynamical graphs, some aspects that go beyond the D-DCSBM assumption should be taken into account.
	
	\medskip
	
	So far, the article dealt with $k=2$ equal-size communities for which the D-DCSBM threshold is well defined. Real networks may of course have multiple asymmetrical-sized classes. As in the static case \cite{bordenave2015non}, we argue that, under this general D-DCSBM setting and the classical assumption that the expected degree of each node is class-independent, the left edge of the bulk spectrum of $H_{\lambda_d,\eta}$ is still asymptotically close to zero and that some of the eigenvectors associated with the isolated negative eigenvalues carry information for community reconstruction.\footnote{In passing, while $\alpha_c(T,\eta)$ is well defined for all $k\geq 2$, when $k>2$, its value no longer corresponds to the position of a detectability threshold, the very notion of which remains an open riddle for $k>2$.} The value $k$ is, in practice, also likely unknown. This also does not affect the idea of the algorithm which exploits all eigenvectors associated to the negative eigenvalues of $H_{\lambda_d,\eta}$, without the need of knowing $k$. The very choice of $k$ is only required by k-means in the last step of spectral clustering and may be performed using off-the-shelf k-means compliant tools, \emph{e.g.}, the \emph{silhouettes method} \cite{rousseeuw1987silhouettes}.
	
	\medskip
	
	Another aspect of practical concern is that successive realizations of $A^{(t)}$ may not be independent across time. Appendix~\ref{app:memory}, covers this issue by introducing \emph{edge persistence} in the model. As suggested in \cite{barucca2018disentangling}, by simply removing from $A^{(t+1)}$ all edges also present in $A^{(t)}$, one then retrieves a sequence of adjacency matrices which, for sparsity reasons, (asymptotically) mimic graphs without edge dependence. These updated adjacency matrices are a suited input replacement to the algorithm.
	
		\setcounter{algorithm}{0}
	
	\begin{algorithm}[t!]
		\begin{algorithmic}[1]
			\State \textbf{Input} : adjacency matrices $\{A^{(t)}\}_{t=1,\dots,T}$ of the undirected graphs $\{\mathcal{G}_t\}_{t = 1,\dots,T}$; label persistence, $\eta$; number of clusters $k$.
			\For {$t = 1:T-1$}
			\State Remove from $A^{(t+1)}$ the edges appearing in both $A^{(t)}$ and $A^{(t+1)}$ (Appendix \ref{app:memory})
			\EndFor
			\State Compute: $d_i^{(t)} \leftarrow \sum_{j=1}^n A_{ij}^{(t)}$;  $c \leftarrow \frac{1}{nT}\sum_{t=1}^T\sum_{i=1}^n d_i^{(t)}$; $\Phi \leftarrow \frac{1}{nTc^2}\sum_{t=1}^T\sum_{i=1}^n \left(d_i^{(t)}\right)^2$; $\alpha_c(T,\eta)$ from Equation \eqref{eq:transition};  $\lambda_d \leftarrow \frac{\alpha_c(T,\eta)}{\sqrt{c\Phi}}$ .
			\State Stack the $m$ eigenvectors of $H_{\lambda_d,\eta}$ with negative eigenvalues in the columns of $X \in \mathbb{R}^{nT\times m}$
			\State Normalize the rows of $X_{i,:} \leftarrow X_{i,:}/{\Vert X_{i,:} \Vert}$
			\For{$t =1:T$}
			\State Estimate the community labels $\{\hat{\ell}_{i_t}\}_{i = 1,\dots n}$ using ${k}$-class \emph{k-means} on the rows $\{X_{i_t}\}_{i = 1,\dots,n}$.
			\EndFor
			\\
			\Return Estimated label vector $\hat{\bm{\ell}} \in \{1,\dots,{k}\}^{nT}$.
			\caption{Community detection in sparse, heterogeneous and dynamical graphs}
			\label{alg:0}
		\end{algorithmic}
	\end{algorithm}

    \medskip

	A last important remark is that $\eta$ is an input of Algorithm~\ref{alg:0}. If unknown, as it would in general be, one may choose an arbitrary $h\in[0,1)$ and $\xi = \alpha_c(T,h)$, to then perform spectral clustering on $H_{\xi,h}$: the leftmost edge of the bulk spectrum of $H_{\xi,h}$ is asymptotically close to zero for all $h$ and consequently Algorithm~\ref{alg:0} can be used in the same form. However, for a mismatched $h$, the detectability threshold now occurs beyond the optimal $\alpha_c(T,\eta)$. Close to the transition, this
	mismatch would give rise to fewer informative isolated negative eigenvalues than expected, resulting in a poor quality label assignment. As a workaround, one may browse through a discrete set of values for $h$ and extract the $h$ maximizing some quality measure, such as the resulting clustering \emph{modularity}. \cite{newman2006modularity}.

    \noindent\textbf{Computation complexity.} The bottleneck of Algorithm~\ref{alg:0} is to compute the embedding $X$. The number of negative eigenvalues $m$ is not \emph{a priori} known and only suspected to be in the interval $\{k,\ldots,kT\}$. Our strategy is to compute the first $k+1$ eigenvectors, ensure that the associated eigenvalues are all negative, then compute the $(k+2)$-th eigenvector, etc., until the largest uncovered eigenvalue crosses zero. This strategy, via standard sparse numerical algebra tools based on Krylov subspaces~\cite{saad_numerical_2011}, costs $\mathcal{O}(nT\sum_{l=k}^m l^2)$. In the best-case (resp., worst-case) scenario, $m=k$ (resp., $m=kT$): the complexity of Algorithm~\ref{alg:0} thus scales as $\mathcal{O}(nTk^2)$ (resp., $\mathcal{O}(nT^4k^3)$).

    \noindent\textbf{An accelerated approximate implementation. }
	As $T$ or $k$ increase, the above complexity may become prohibitive. A recent workaround strategy~\cite{tremblay_icassp16, ramasamy_compressive_2015, tremblay_icml16}, based on polynomial approximation and random projections, is here particularly adapted, and decreases the overall complexity of the algorithm to $\mathcal{O}(nTk\log(nT))$, for a limited loss in precision. The resulting fast implementation is described in Algorithm~\ref{alg:fast} and detailed in Appendix~\ref{app:fast_algo}. To give an order of magnitude, a simulation\footnote{The laptop's RAM is $7.7$ Gb with Intel Core i7-6600U CPU @ 2.6GHz x 4.} of Algorithm~\ref{alg:0} for $n = 10^5$, $T=5$ (resp., $n=5\,000$, $T=100$), $k=2$, $c=6$, $\eta=0.5$, $\Phi = 1.6$, $\alpha=2\alpha_c(T,\eta)$ takes on average approximately $1$~minute (resp., $40$~minutes), whereas Algorithm~\ref{alg:fast} converges in less than $4$~minutes in both cases. The reader is referred to Appendix~\ref{app:fast_algo} for more details.

	\begin{figure}[t!]
		\centering
		\includegraphics[width = \columnwidth]{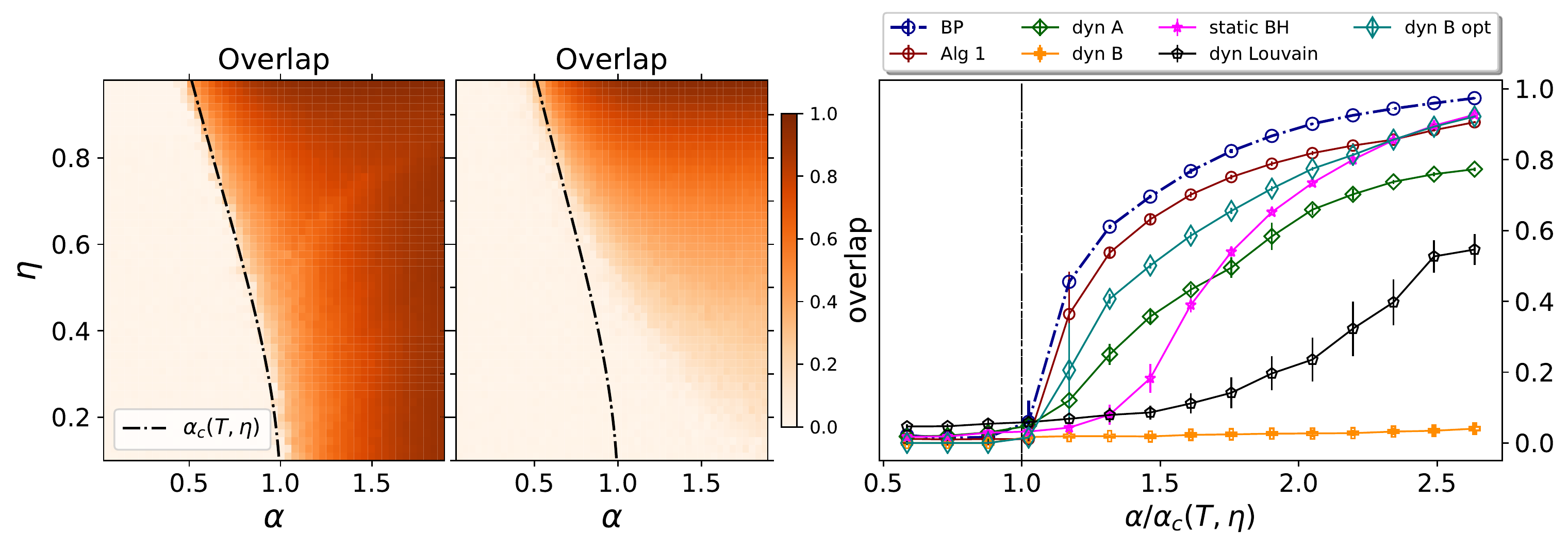}
		\caption{\textbf{Left}: overlap comparison at $t = T$ for Algorithm \ref{alg:0} vs.\@ \cite{keriven2020sparse}, in color gradient, for various detectability hardness levels $\alpha$ ($x$-axis) and label persistence $\eta$ ($y$-axis); $n = 10\,000$, $T=5$, $c = 10$, $\Phi = 1$; averaged over $4$ samples. \textbf{Right}: mean overlap across all values of $t$, as a function of $\alpha$, for Algorithm~\ref{alg:0} (Alg~1), BP \cite{ghasemian2016detectability}, the dynamic adjacency matrix of \cite{keriven2020sparse} (dyn~A), the dynamical non-backtracking of \cite{ghasemian2016detectability} (dyn~B and dyn~B opt), the static Bethe-Hessian of \cite{dall2019revisiting} (static BH) and the dynamical Louvain algorithm of \cite{mucha2010community} (dyn~Louvain); $n=5\,000$, $T=4$, $c=6$, $\eta=0.7$, $\Phi=1$; averaged over $20$ samples ($3$ for BP). For all plots, $k=2$.}
		\label{fig:ov_vs_a}
	\end{figure}

	\subsection{Performance comparison on synthetic datasets}

	Figure~\ref{fig:ov_vs_a} shows the performance of different clustering algorithms in terms of overlap
	\begin{equation}
	{\rm ov}(\bm{\ell},\bm{\hat{\ell}}) = \underrel{\bm{\bar{\ell}} \in \mathcal{P}(\bm{\hat{\ell}})}{\rm max} \frac{1}{1-\frac{1}{k}}\left(\frac{1}{n}\sum_{i=1}^n \mathds{1}_{\ell_i,\bar{\ell}_i} -\frac{1}{k}\right),
	\label{eq:overlap}
	\end{equation}	
	where $\bm{\ell}, \bm{\hat{\ell}} \in \{1,\dots,k\}^n$ are the ground truth and estimated label vectors, respectively, while $\mathcal{P}(\bm{\ell})$ is the set of permutations of $\bm{\ell}$. The overlap ranges from zero for a random label assignment to one for perfect label assignment. Figure~\ref{fig:ov_vs_a}-left compares the overlap performance as a function of $\alpha$ and $\eta$ for Algorithm~\ref{alg:0} versus the adjacency averaging method of \cite{keriven2020sparse} (which we recall assumes $\eta=1-o_n(1)$). The overlap is only considered at $t=T$ so to compare Algorithm~\ref{alg:0} on even grounds with \cite{keriven2020sparse} which only outputs one partition (rather than one for every $t$). The theoretical detectability threshold line $\alpha=\alpha_c(T,\eta)$ visually confirms the ability of Algorithm~\ref{alg:0} to assign non trivial class labels as soon as theoretically possible, as opposed to the method of \cite{keriven2020sparse} which severely fails at small values of $\eta$. 
	
	Figure~\ref{fig:ov_vs_a}-right then compares the average overlap performance of Algorithm~\ref{alg:0} against competing methods, for varying detection complexities $\alpha/\alpha_c(T,\eta)$. Algorithm~\ref{alg:0} is outperformed only by the BP algorithm\footnote{The codes used to obtain the BP performance displayed in Figure~\ref{fig:ov_vs_a} are courtesy of Amir Ghasemian.}, but has an approximate $500$-fold reduced computational cost. 
    The  computational heaviness of BP becomes practically prohibitive for larger values of $n$. For completeness, Appendix~\ref{app:simulation} provides further numerical performance comparison tests for different values of $\eta$, $\Phi$, for $k > 2$, for larger values of $n$ and $T$, and for graphs with clusters of different average sizes. Interestingly, for large values of $\alpha$, Algorithm~\ref{alg:0} is slightly outperformed by the static Bethe-Hessian of \cite{dall2019revisiting}, independently run at each time-step. As discussed at the end of Section~\ref{sec:main},  the choice $\xi = \lambda_d$ is sub-optimal compared to the optimal (but out-of-reach in practice) choice $\xi=\lambda$, the difference becoming more visible as $\alpha$ increases away from $\alpha_c$. Supposing one has access to an oracle for $\lambda$, running Algorithm~\ref{alg:0} on $H_{\lambda,\eta}$ outputs a performance in terms of overlap (not shown) that is first super-imposed with the ``Alg 1'' plot for small values of $\alpha$ and gradually converges to the performance of ``BP'' as $\alpha$ increases; thus outperforming ``static BH'' everywhere.
    From a dynamical viewpoint, also, the large $\alpha$ regime is of least importance as a static algorithm can, alone, output a perfect reconstruction. Further numerical experiments are shown in Appendix~\ref{app:simulation}.
    
    For the non-backtracking method of \cite{ghasemian2016detectability} (``dyn B''), the authors suggest to use (as we did here) the eigenvector associated to the second largest eigenvalue of $B_{\lambda,\eta}$, which, as $H_{\lambda_d,\eta}$,  may also have informative and uninformative eigenvalues in reversed order. The curve ``dyn B opt" shows the performance obtained using all the isolated eigenvectors of $B_{\lambda,\eta}$ and it confirms -- in agreement with Appendices~\ref{app:proof} and \ref{app:simulation} and the claims of \cite{ghasemian2016detectability} -- that $B_{\lambda,\eta}$ can indeed make non-trivial community reconstruction for all $\alpha > \alpha_c(T,\eta)$. Note that, as in the static case \cite{krzakala2013spectral, saade2014spectral}, $B_{\lambda,\eta}$ is outperformed by $H_{\lambda_d,\eta}$ which, additionally, is symmetric and smaller in size, is well defined regardless of $\lambda$ and is, therefore, a more suitable candidate for community detection.

\subsection{Test on Sociopatterns \emph{Primary school}}
\label{sec:sociopatterns}
    
\begin{figure}[t!]
	\centering
	\includegraphics[width=0.8\columnwidth]{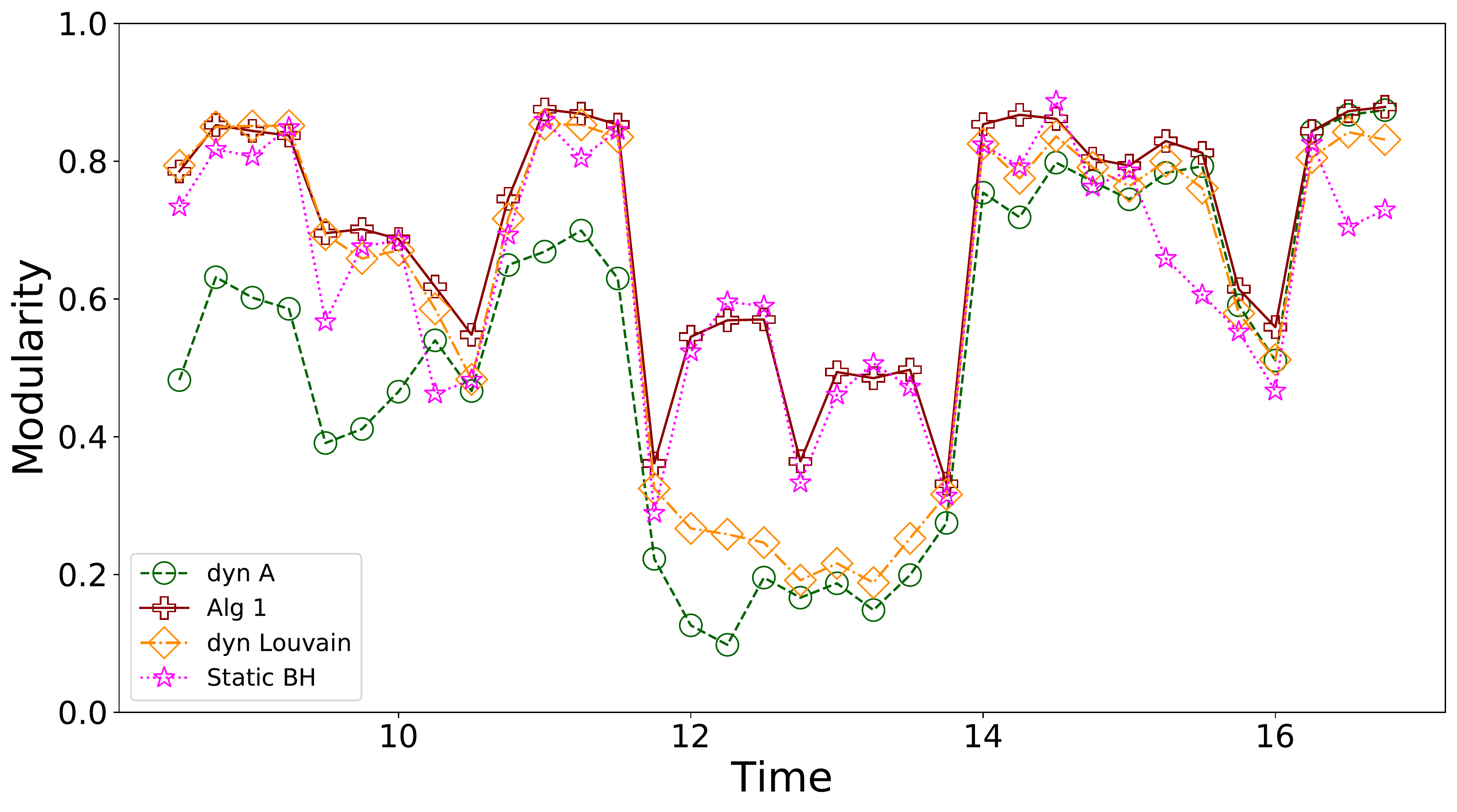}
	\caption{Modularity as a function of time for Algorithm~\ref{alg:0} (Alg 1) for $\eta = 0.55$, the dynamic adjacency matrix of \cite{keriven2020sparse} (dyn A), the dynamic Louvain algorithm \cite{mucha2010community} (dyn Louvain) and the static Bethe-Hessian of \cite{dall2019revisiting} (static BH). The graph $\{\mathcal{G}_t\}_{t = 1,\dots,T}$ is obtained from the \emph{Primary school} network \cite{gemmetto2014mitigation,stehle2011high} dataset, as in Section~\ref{sec:sociopatterns}. For Algorithm~\ref{alg:0}, \cite{dall2019revisiting} and \cite{keriven2020sparse}, $k=10$ is imposed.}
	\label{fig:real}
\end{figure}

This section shows the results of our experiments on the \emph{Primary school} network \cite{gemmetto2014mitigation,stehle2011high}  of the  \href{www.sociopatterns.org}{SocioPatterns} project. The dataset contains a temporal series of contacts between children and teachers of ten classes of a primary school. For each time $1 \leq t \leq T$, $\mathcal{G}_t$ is obtained considering all interactions from time $t$ to time $t + 15~{\rm min}$, starting from $t_1 =$ 8:30 am until $t_T =$ 5 pm for $T = 33$. Figure~\ref{fig:real} compares the modularity as a function of time for different clustering techniques. We empirically observe that, for this dataset, multiple values of $\eta$ give similar results: this is not surprising because the clusters are here well delineated and we are in the (less interesting) easy detection regime. The value $\eta = 0.55$ is considered as an input of Algorithm~\ref{alg:0}, because it approximately matches the value of $\eta$ estimated from the inferred label vector $\bm{\hat{\ell}}$ (see Equation~\eqref{eq:label_markov}).

Figure~\ref{fig:real} shows that Algorithm~\ref{alg:0} is better than \cite{keriven2020sparse,mucha2010community} at all times, with a drastic gain during the lunch break, in which the community structure is harder to delineate. As compared to the static Bethe-Hessian, Algorithm~\ref{alg:0} is slightly outperformed only on some times during the lunch break, while for other times it benefits from the positive correlation of the labels. 
Defining a unique, time independent $\eta$ certainly hampers the performance on this specific dataset in which a very large $\eta$ is expected during the lesson times, while a small $\eta$ may be more appropriate during the lunch break.

	\section{Concluding remarks}
	\label{sec:conclusion}
	
	By means of arguments at the crossroads between statistical physics and graph theory, this article tailored Algorithm~\ref{alg:0}, a new spectral algorithm for community detection on sparse dynamical graphs. Algorithm~\ref{alg:0} is capable of reconstructing communities as soon as theoretically possible, thereby largely outperforming state-of-the-art competing spectral approaches (especially when classes have a short-term persistence) while only marginally under-performing the (theoretically claimed optimal but computationally intensive) belief propagation algorithm.

	A delicate feature of Algorithm~\ref{alg:0} concerns the estimation of the class-persistence parameter $\eta$, if not available. We hinted in Section~\ref{sec:algo} at a greedy line-search solution which is however computationally inefficient and lacks of a sound theoretical support. This needs be addressed for Algorithm~\ref{alg:0} to be more self-contained and applicable to the broadest range of practical networks.

	Beyond this technical detail, the present analysis only scratches the surface of dynamical community detection: the problem in itself is vast and many degrees of freedom have not been here accounted for. The label persistence $\eta$ and community strength matrix $C$ (and thus the parameter $\lambda$ in a symmetric two-class setting) are likely to evolve with time as well. We empirically observed that Algorithm~\ref{alg:0} naturally extends to this setting, each temporal block of the matrix $H_{\cdot,\cdot}$ now using its corresponding $\lambda_d^{(t)}$ and $\eta_t$. Yet, while Algorithm~\ref{alg:0} seems resilient to a more advanced dynamical framework, the very concept of \emph{detectability thresholds} becomes more elusive in a symmetrical two-class setting: a proper metric to measure the distance to optimality would thus need to be first delineated.
	
	\include{supp}

	\section*{Broader impact}
	
	Community detection algorithms have a broad interest as they can be applied to a very vast class of problems and settings. An interesting example, of utmost importance in the present days, was given by \cite{gauvin2015revealing} were the authors showed the importance of keeping track of the time-evolving community structure of social networks to properly model an epidemic spreading. Not unlike any other clustering algorithm, however, when applied to a real social network, our algorithm can potentially evidence differences in terms of \emph{e.g.} race, sex, religion. As discussed in \cite{sandvig2016automation}, if such an output is used in some decision process, the result can indeed produce discriminatory choices.
	
	Although we are aware of the potential weaknesses, the mainly theoretical nature of our study, as well as the nowadays vast literature in the field of community detection, allows us to not foresee any major negative consequence from our study. On the contrary, keeping into account of the realistic time-evolving nature of networks can allow to improve and better understand our studies in the field.
	
 	\section*{Acknowledgements}

RC's work is supported by the MIAI LargeDATA Chair at University Grenoble-Alpes and the GIPSA-HUAWEI Labs project Lardist. NT's work is partly supported by the French National Research Agency in the framework of the "Investissements d’avenir” program (ANR-15-IDEX-02) and the LabEx PERSYVAL (ANR-11-LABX-0025-01).

\end{document}

%% file: supp.tex
%\renewcommand{\thesubsection}{\Alph{subsection}}
%\def\theequation{\thesection.\arabic{equation}}
%\numberwithin{equation}{subsection}

%\renewcommand{\thesection}{\Alph{section}}
%\def\theequation{\thesection.\arabic{equation}}
%\numberwithin{equation}{section}

\appendix

\section*{\Large Supplementary material}

The supplementary material provides complementary technical arguments to the main results of the article (Sections~\ref{app:detectability}--\ref{app:proof}), along with a discussion on the extension of the present setting to dynamic graphs with \emph{link persistence} across time (Section~\ref{app:memory}). Further numerical tests on the performance of Algorithm~\ref{alg:0} are presented in Section~\ref{app:simulation}, while Section~\ref{app:fast_algo} presents the detailed description of Algorithm~\ref{alg:fast} to handle fast approximate spectral clustering.

\section{Detectability threshold for finite $T$}
\label{app:detectability}
This section discusses the conjecture of \cite{ghasemian2016detectability} in which the authors introduce a threshold $\alpha_c(T,\eta)$ (however not explicitly defined\footnote{Precisely, quoting the authors, this is as far as $\alpha_c(T,\eta)$ is defined: ``We can compute the corresponding finite-time threshold for a fixed $T$ by diagonalizing a $(3T-2)$-dimensional matrix, where we have a branching process with states corresponding to moving along spatial, forward-temporal, or backward-temporal edges at each time step''.}), below which ($\alpha < \alpha_c(T,\eta)$) community detection is not feasible. We go here beyond \cite{ghasemian2016detectability} by providing an explicit value for $\alpha_c(T,\eta)$ for all finite $T$.

As a consequence of the sparsity of each $\mathcal{G}_t$, the graph $\mathcal{G}$, obtained by connecting together the same node at successive times as per Definition~\ref{def:G} (recall Figure~\ref{fig:G}) is locally tree-like, \emph{i.e.} the local structure of $\mathcal{G}$ around a node $v \in \mathcal{V}$ is the same as that of a Galton-Watson tree $\mathcal{T}(v)$ \cite{dembo2010gibbs}, rooted at $v$, designed according to the following procedure: let $\ell_{v} \in \{1,2\}$ be the label of $v$; next generate its progeny by creating $d_s$ spatial children (\emph{i.e.}, nodes which live at the same time as $v$), where $d_s$ is a Bernoulli random variable with mean $c\Phi$, and two temporal children (\emph{i.e.}, nodes which are the projection of $v$ at neighbouring times); for each spatial child $w$, assign the label $\ell_w = \ell_{v}$ with probability $c_{\rm in}/(c_{\rm in}+c_{\rm out})$ and $\ell_w = 3-\ell_{v}$ otherwise; the temporal children keep the same label as $v$ with probability $(1+\eta)/2$ and change it with probability $(1-\eta)/2$; each node thus created further generates its own set of offspring, with the only difference that the temporal children only bear one extra temporal child, while spatial children bear two.

In the limit $n, T \to \infty$, for any arbitrary $v \in \mathcal{V}$, the local structure of $\mathcal{G}$ around $v$ is the same as $\mathcal{T}(v)$, the Galton-Watson tree rooted at $v$. This means that, within a neighborhood reachable in a finite number of steps from $v$ in $\mathcal{G}$ or $\mathcal{T}(v)$, the probability distribution of the labels is asymptotically the same. The local tree-like structure is preserved for finite $T$ (and $n \to \infty$) but the boundary conditions imposed by $t = 1$ and $t = T$ must be accounted for.

This said, in \cite{janson2004robust}, the authors show that, for a Galton-Watson tree in which only spatial children are present, label reconstruction is feasible if and only if $c\Phi\lambda^2 = \alpha^2 > 1 = \alpha_c^2$, where $\lambda = (c_{\rm in} - c_{\rm out})/(c_{\rm in} + c_{\rm out})$. In \cite{ghasemian2016detectability}, the authors conjectured a generalization of this result for a multi-type branching process, such as just described to construct $\mathcal{T}(v)$. In this setting, each node acts differently depending on its being a spatial or a temporal child. In the former case, two temporal children are generated (with label covariance equal to $\eta$), while in the latter only one temporal child is generated. The conjecture of \cite{ghasemian2016detectability} (which we adapted to the D-DCSBM) states that, for $T\to \infty$, community detection is possible if and only if the largest eigenvalue of
\begin{align}
M_{\infty}(\alpha,\eta) = 
\begin{pmatrix}
\alpha^2 &2\eta^2 \\
\alpha^2 & \eta^2
\end{pmatrix}
\label{eq:M_infty}
\end{align}
is greater than one. This condition is verified as long as $\alpha > \alpha_c(\infty,\eta) = \sqrt{(1-\eta^2)/(1+\eta^2)}$. 

The authors of \cite{ghasemian2016detectability} also provided directions to extend their result to finite $T$, which we here make explicit. For each time instant, three types of edges exist: spatial edges (connecting nodes in $\mathcal{G}_t$ to nodes in $\mathcal{G}_t$), forward temporal edges (connecting nodes in $\mathcal{G}_t$ to nodes in $\mathcal{G}_{t+1}$) and backwards temporal edges (connecting nodes in $\mathcal{G}_t$ to nodes in $\mathcal{G}_{t-1}$).
We then construct a matrix $\tilde{M}_T(\alpha,\eta) \in \mathbb{R}^{3T\times 3T}$ identifying the rows and the columns as $\{({\rm backwards~temporal})_t, ({\rm spatial})_t, ({\rm forward~temporal})_t\}_{t=1,\dots,T}$. A $({\rm backwards~temporal})_t$ edge goes from a node in $\mathcal{V}_t$ to a node in $\mathcal{V}_{t-1}$ that has, on average, $c\Phi$ spatial children with label correlation equal to $\lambda$ and one backwards temporal child, with label correlation equal to $\eta$. Similarly $({\rm spatial})_t$ goes from a node in $\mathcal{V}_t$ to a node in $\mathcal{V}_t$ having $c\Phi$ temporal children and, one forward and one backwards temporal children; finally, $({\rm forward~temporal})_t$ goes from $\mathcal{V}_t$ to $\mathcal{V}_{t+1}$ with one forward temporal child and $c\Phi$ spatial children. The entry $i,j$ of $
\tilde{M}_T(\alpha,\eta)$ is then set equal to the number of off-springs of type $j$ of a node reached by an edge of type $i$, multiplied by the square label correlation. As forward temporal edges do not exist for $t=T$ and backwards temporal edges do not exist for $t = 1$, the matrix $\tilde M_T(\alpha,\eta) \in \mathbb{R}^{3T\times 3T}$ takes the form
\begin{align}
\tilde M_T(\alpha,\eta) = \begin{pmatrix}
\tilde{M}_d^+ & M_+ & 0  & \dots&0 \\
\tilde{M}_- & M_d & \ddots &\dots &0 \\
0&M_{-}&\ddots&M_+&0&\\
\vdots&\vdots&\ddots&M_d&\tilde{M}_+\\
0&0&\dots& M_-&\tilde{M}_d^-
\end{pmatrix}
\label{eq:transition2}
\end{align}
where 
\begin{align}
&M_d = \begin{pmatrix}
0&0&0 \\
\eta^2 & c\Phi\lambda^2 & \eta^2 \\
0&0&0
\end{pmatrix}; \quad
M_+ = \begin{pmatrix}
0&0&0 \\
0&0&0 \\
0 & c\Phi\lambda^2 & \eta^2
\end{pmatrix}; \quad
M_- = \begin{pmatrix}
\eta^2&c\Phi\lambda^2&0 \\
0&0&0 \\
0&0&0
\end{pmatrix} \nonumber\\
&
\tilde{M}_d ^+= \begin{pmatrix}
0&0&0 \\
0 & c\Phi\lambda^2 & \eta^2 \\
0&0&0
\end{pmatrix}; \quad
\tilde{M}_+ = \begin{pmatrix}
0&0&0 \\
0&0&0 \\
0 & c\Phi\lambda^2 & 0
\end{pmatrix}; \nonumber \\
&
\tilde{M}_- = \begin{pmatrix}
0&c\Phi\lambda^2&0 \\
0&0&0 \\
0&0&0
\end{pmatrix}; \quad \tilde{M}_d ^-= \begin{pmatrix}
0&0&0 \\
\eta^2 & c\Phi\lambda^2 & 0 \\
0&0&0
\end{pmatrix}.  \nonumber
\end{align}

\begin{figure}[t!]
	\centering
	\includegraphics[width = .9\columnwidth]{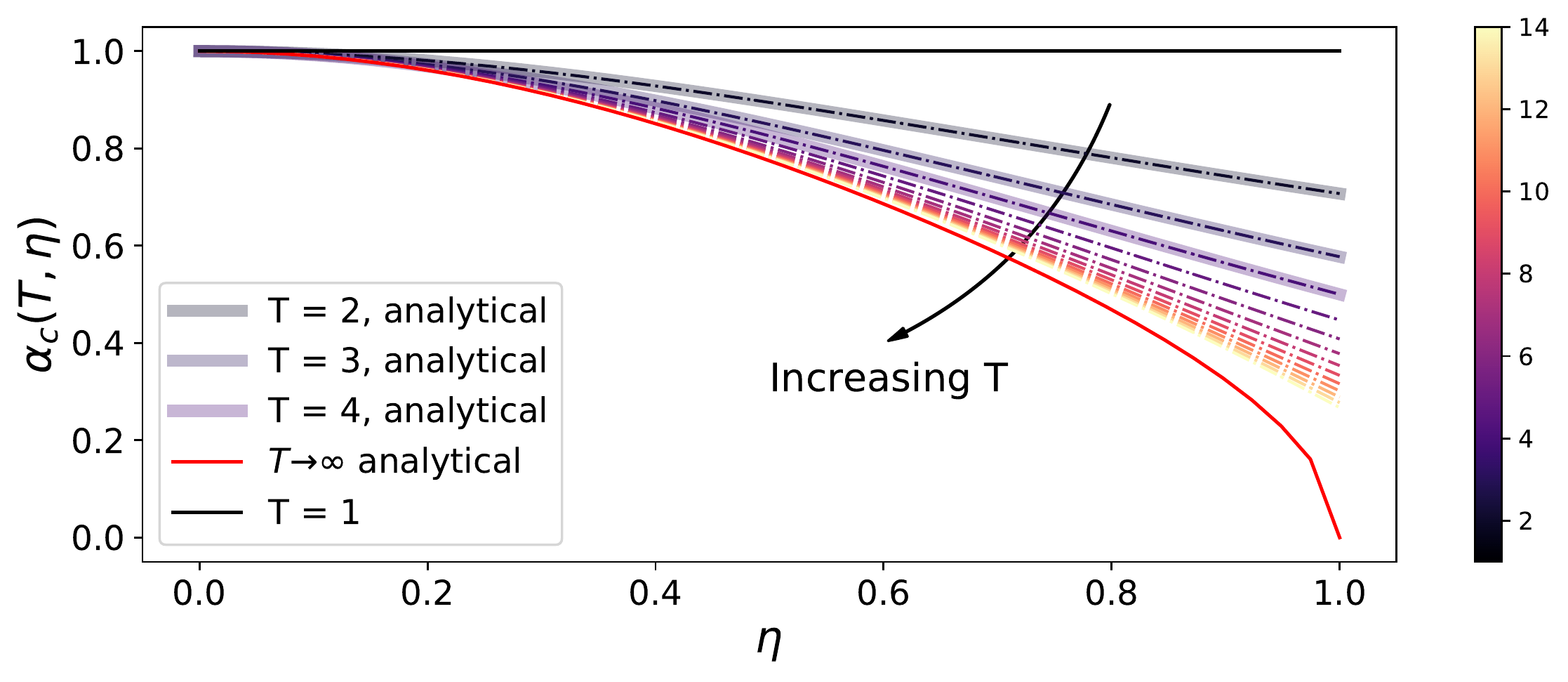}
	\caption{Position of $\alpha_c(T,\eta)$ as a function of $\eta$ (x axis) and $T$ (color code). The black and red solid lines correspond to the analytical values of $\alpha_c(T,\eta)$, for $T=1$ and $T=\infty$, respectively. The dashed dotted lines are the position of $\alpha_c(T,\eta)$ computed numerically, and the thick solid pale lines are the analytical values of $\alpha_c(T,\eta)$ for $T\in\{2,3,4\}$.}
	\label{fig:transition}
\end{figure}

Note that, since the first and the last rows of $\tilde{M}_T(\alpha,\eta)$ only have zero entries, $\tilde{M}_T(\alpha,\eta)$ has the same non-zero eigenvalues as $M_T(\alpha,\eta)$ defined in Equation~\eqref{eq:matrix_M}. This also implies that $M_T(\alpha,\eta)$ shares the non-zero eigenvalues of a matrix of size $(3T-2)\times (3T-2)$ as initially conjectured in \cite{ghasemian2016detectability}.

The analytical expression of $\alpha_c(T,\eta)$ can be obtained for $T=2,3,4$ and is reported in the main text. For all other values of $T$ it can be computed numerically. The value of $\alpha_c(T,\eta)$ as a function of $\eta$ is reported in Figure~\ref{fig:transition} for different values of $T$.

\section{Derivation of the dynamic Bethe-Hessian matrix}
\label{app:BH}

This appendix derives the matrix $H_{\xi,h}$, which arises from the variational Bethe approximation applied to the Hamiltonian of Equation~\eqref{eq:ising}, which we recall assumes the form
\begin{align}
\mathcal{H}_{\xi,h}(\bm{s}) = -\sum_{t = 1}^T\left( \sum_{(i_t,j_t)\in\mathcal{E}_t}{\rm ath}(\xi) s_{i_t}s_{j_t} + \sum_{i_t \in \mathcal{V}_t}{\rm ath}(h) s_{i_t}s_{i_{t+1}}\right).
\end{align}
Collecting all time instants, $\mathcal{H}_{\xi,h}(\bm{s})$ can be synthetically written under the form
\begin{align}
\mathcal{H}_{\xi,h}(\bm{s}) = -\sum_{(ij) \in \mathcal{E}} {\rm ath}(\omega_{ij})~s_is_j
\end{align}
for some appropriate coupling $\omega_{ij}$ (and where we recall that $\mathcal{E}$ is the set of all edges of $\mathcal{G}$). Each realization $\bm{s}$ is a random variable, drawn from the Maxwell-Boltzmann distribution
\begin{align}
\mu(\bm{s}) = \frac{1}{Z}e^{-\mathcal{H}_{\xi,h}(\bm{s})},
\label{eq:boltzmann}
\end{align}
where $Z$ is the normalization constant. We are interested in the average realization of $\bm{s}$ over the distribution $\mu(\cdot)$, that we denote $\bm{m}^* = \langle \bm{s}\rangle$, with $\langle\cdot\rangle$ being the average over \eqref{eq:boltzmann} . From Equation~\eqref{eq:boltzmann}, configurations having a small energetic cost will occur with a larger probability but there are very few such configurations, as opposed to the exponentially many disordered ones. The competing behavior of these two terms defines two regimes: (i) the small interaction regime (called the \emph{paramagnetic phase}, for small $\xi$ and $h$) in which the disordered configurations dominate the average configuration (which is the null vector) and (ii) the strong interaction regime (for large $\xi$ and $h$) in which the average value of $\bm{s}$ is non-trivial and is dominated by the \emph{modes} of $\bm{s}$ which are local minima of the Hamiltonian of Equation~\eqref{eq:ising}. These modes are determined by the ``mesoscale'' structure of $\mathcal{G}$. 

 \medskip 
 
The value of $\bm{m}^*$ cannot be computed exactly but, given the locally tree-like nature of $\mathcal{G}$, it may be evaluated using the asymptotically exact variational Bethe approximation \cite{yedidia2003understanding}. This approximation $P_{\rm Bethe}(\cdot)$ of $\mu(\cdot)$ reads
\begin{align}
{P}_{\rm Bethe}(\bm{s}) = \frac{\prod_{(i,j)\in\mathcal{E}}P_{ij}(s_{i}s_{j})}{\prod_{i\in\mathcal{V}}[P_{i}(s_{i})]^{d_{i}-1}},
\label{eq:PBethe}
\end{align} 
where $P_{ij}(\cdot)$ and $P_i(\cdot)$ are the edge and node marginals of ${P}_{\rm Bethe}$ and $d_i$ is the \emph{total degree on $\mathcal{G}$} of node $i$. Further defining the free energy and the Bethe free energy respectively as
\begin{align}
F &= \sum_{\bm{s}} \mu(\bm{s})\left[\mathcal{H}_{\xi,h}(\bm{s}) +{\rm log}~\mu(\bm{s})\right] = -{\rm log}~Z \\
F_{\rm Bethe}(\bm{m},\bm{\chi}) &= \sum_{\bm{s}} P_{\rm Bethe}(\bm{s})\left[\mathcal{H}_{\xi,h}(\bm{s}) +{\rm log}~P_{\rm Bethe}(\bm{s})\right],
\label{eq:FBethe}
\end{align}
where $m_i = \langle\sigma_i\rangle_{\rm Bethe}$ and $\chi_{ij} = \langle\sigma_i\sigma_j\rangle_{\rm Bethe}$, $\langle \cdot \rangle_{\rm Bethe}$ denoting the average taken over $P_{\rm Bethe}(\cdot)$. From a direct calculation, it comes that $F_{\rm Bethe}(\bm{m},\bm{\chi})-F = D_{\rm KL}(P_{\rm Bethe}||\mu) \geq 0$, where $D_{\rm KL}(\cdot)$ is the Kullback-Leibler divergence. Therefore, by minimizing $F_{\rm Bethe}$ with respect to $\bm{m}$, one minimizes the divergence with respect to the real distribution and obtains an optimal estimate for $\bm{m}^*$. 

The Bethe free energy can be obtained by plugging Equation~\eqref{eq:PBethe} into Equation~\eqref{eq:FBethe} and takes the explicit form
\begin{align}
F_{\rm Bethe}(\bm{m},\bm{\chi}) =& -\sum_{(ij)\in\mathcal{E}}{\rm ath}(\omega_{ij})~\chi_{ij}+\sum_{(ij)\in\mathcal{E}}\sum_{s_{i}s_{j}}f\left(\frac{1+m_{i}s_{i}+m_{j}s_{j}+\chi_{ij}s_{i}s_{j}}{4}\right) \nonumber \\
&- \sum_{i \in \mathcal{V}} (d_{i}-1)\sum_{s_{i}}f\left(\frac{1+m_{i}s_{i}}{2}\right),
\end{align}
where $f(x) = x{\rm log}(x)$. In the case of weak interactions (small $\omega_{ij}$), $F_{\rm Bethe}$ has a unique minimum in $\bm{m} = 0$. For larger values of $\omega_{ij}$, it has a global minimum at $\bm{m} \propto \bm{1}_{nT}$ and other local minima appear, corresponding to configurations correlated with the mesoscale structure of $\mathcal{G}$. In order to study along which directions the function $F_{\rm Bethe}$ finds its local minima, one needs to evaluate the Hessian matrix of $F_{\rm Bethe}$ at $\bm{m} = 0$, as done in \cite{watanabe2009graph,saade2014spectral}, to obtain
\begin{align}
\left.\frac{\partial^2 F_{\rm Bethe}(\bm{m},\bm{\chi})}{\partial{m_i}\partial{m_j}}\right|_{\bm{m} = 0} = -\frac{\chi_{ij}}{1-\chi_{ij}^2}A_{ij} + \left(\sum_{k \in \partial i} \frac{1}{1-\chi_{ik}^2} -(d_i-1)\right)\mathds{1}_{ij}~,
\end{align}
where $A \in \{0,1\}^{nT}$ is the adjacency matrix of $\mathcal{G}$ and $d_i = [A\bm{1}_n]_i$. Similarly minimizing $F_{\rm Bethe}$ with respect to $\chi_{ij}$,
\begin{align}
\left.\frac{\partial F_{\rm Bethe}(\bm{m},\bm{\chi})}{\partial \chi_{ij}}\right|_{\bm{m} = 0} = -{\rm ath}(\omega_{ij}) + {\rm ath}(\chi_{ij}) = 0
\end{align}
and so $\chi_{ij} =  \omega_{ij}$. 

To finally retrieve the expression of Equation~\eqref{eq:BHMatrix}, note that $d_i = d_i^{(t)} + 2$ if $2 \leq t \leq T-1$ and $d_i = d_i^{(t)} + 1$ otherwise, where $d_i^{(t)}$ is the degree of node $i$ in $\mathcal{G}_t$, and impose
\begin{align}
\chi_{ij} = \begin{cases}
\xi \quad &{\rm if}~\exists~t~{\rm such~that~}i,j \in \mathcal{V}_t \\
h \quad &{\rm otherwise}
\end{cases}
\end{align}
as requested.

We therefore retrieve the matrix $H_{\xi,h}$ of Equation~\eqref{eq:BHMatrix}. When $H_{\xi,h}$ has a negative eigenvalue, $\bm{m} = 0$ is a saddle point and the free energy has a local minimum for some non-trivial configuration. The eigenvector associated to this negative eigenvalue points towards the direction of the local minimum of $F_{\rm Bethe}$. As discussed in Section~\ref{sec:main}, the directions along which stable configurations are observed correspond to the dominant modes appearing in the Hamiltonian and are naturally correlated to the class structure. The smallest eigenvalue-eigenvector pairs of $H_{\xi, h}$ may thus be used to retrieve information on the directions of the dominant informative modes of the graph, as depicted in Figure~\ref{fig:stable}.

\section{Technical results of Section~\ref{subsec:main.th}}
\label{app:proof}

This section provides theoretical support to Proposition~\ref{prop:2} of Section~\ref{subsec:main.th}. 

\medskip

Exploiting the deep relation -- which we detail in Section~\ref{subapp:BH_B} -- that there is between the dynamical Bethe-Hessian of Equation~\eqref{eq:BHMatrix} and the weighted non-backtracking matrix of Equation~\eqref{eq:matrix_B}, we study the spectrum of the latter to infer some important properties of our proposed dynamical Bethe-Hessian. In particular, the eigenvalues of the non-backtracking matrix can be divided into two groups: (i) a majority of eigenvalues contained in a disc in the complex plane which delimits the \emph{bulk} of this matrix (ii) few isolated eigenvalues with modulus larger than the radius of the bulk. These properties are known and well established in the static regime \cite{bordenave2015non, gulikers2016non} and we empirically observed to be maintained also in the dynamical setting under study. Furthermore, in the case of $k=2$ classes, in the static case, the isolated eigenvectors (with largest modulus) are the Perron-Frobenius eigenvector (with all positive entries) and the eigenvector useful for community reconstruction. Similarly, in the dynamical case we have two \emph{families} of eigenvectors (see Figure~\ref{fig:stable}) coming from these two modes. We will refer to them as \emph{informative family} and \emph{uninformative family}. 

Based on these empirical observations, we formulate the following assumption:
\begin{assumption}
	Let $\mathcal{G}$ be a graph generated according to Definition~\ref{def:G} and $B_{\xi,h}$ the matrix defined in Equation~\eqref{eq:matrix_B}. The bulk of $B_{\xi,h}$ is bounded by a disk in the complex plane with radius denoted by $L_{\xi,h}$. The eigenvalues with modulus larger than $L_{\xi, h}$ are isolated and their corresponding eigenvector are determined by the mesoscale structure of $\mathcal{G}$. 
	\label{ass:B}
\end{assumption}

Based on this assumption, in Section~\ref{subapp:proof.spectrum} we determine the asymptotic position of the isolated eigenvalues with modulus larger than the radius of the bulk, as well as the radius of the bulk itself; from these results, Sections~\ref{subapp:proof.prop2} concludes on Proposition~\ref{prop:2}. In passing, with the results of \ref{subapp:proof.spectrum}, some properties of the spectrum of the dynamical non-backtracking matrix of \cite{ghasemian2016detectability} are also discussed.

\subsection{Bethe-Hessian and weighted non-backtracking matrices}
\label{subapp:BH_B}

Let us first elaborate on an important property connecting the spectra of the Bethe-Hessian and non-backtracking matrices. This relation is well known in the literature (see \emph{e.g} \cite{saade2016spectral, terras2010zeta, watanabe2009graph}). For sake of clarity, we here report the main results that relate the eigenvalues and eigenvectors of the two matrices. Let us consider the following two matrices for arbitrary weights $\bm{\omega} = \{\omega_{ij}\}_{(ij) \in \mathcal{E}}$ such that $\omega_{ij} < 1$ for all $(ij) \in \mathcal{E}^d$, the set of directed edges of $\mathcal{G}$:
\begin{align}
\left(B_{\bm{\omega}}\right)_{(ij)(k\ell)} &= \mathds{1}_{jk}(1-\mathds{1}_{il})~\omega_{kl} \quad \forall~(ij),(kl) \in \mathcal{E}^d, ~  \label{eq:new_B}\\
\left(H_{\bm{\omega}}\right)_{ij} &= \left(1+\sum_{k \in\partial i}\frac{(\omega_{ik}/x)^2}{1-(\omega_{ik}/x)^2}\right)\mathds{1}_{ij} - \frac{(\omega_{ik}/x)}{1-(\omega_{ik}/x)^2}A_{ij}, \quad \forall~i,j\in \mathcal{V}, ~x \in (1,\infty)
\label{eq:new_H}
\end{align}

We now show that, for $\omega_{ij} < 1$, whenever $x \geq 1$ is a real eigenvalue of $B_{\bm{\omega}}$, $\det[H_{\bm{\omega }/x}]=0$. Indeed, let $\bm{g} \in \mathbb{R}^{|\mathcal{E}^d|}$ be an eigenvector of $B_{\bm{\omega}}$ with eigenvalue $x \geq 1$. Then
\begin{align}
\left(B_{\bm{\omega}}\bm{g}\right)_{ij} = \sum_{k \in \partial j \setminus i} \omega_{jk}g_{jk} = m_j - \omega_{ji}g_{ji} = xg_{ij},
\end{align}
where $m_j \equiv \sum_{k \in \partial j} \omega_{jk}g_{jk}$. We may gather this relation under the system of equations
\begin{align}
\begin{pmatrix}
m_j \\
m_i
\end{pmatrix} = 
\begin{pmatrix}
x & \omega_{ij} \\
\omega_{ij} & x
\end{pmatrix}
\begin{pmatrix}
g_{ij} \\
g_{ji}
\end{pmatrix}.
\end{align}
Since $\omega_{ij}^2 < 1$ for all $(i,j)$, the system is invertible and a straightforward calculation gives
\begin{align}
m_i = \sum_{j \in \partial i} \frac{\omega_{ij}x}{x^2-\omega_{ij}^2}m_j - m_i \sum_{j \in \partial i}\frac{\omega_{ij}^2}{x^2-\omega_{ij}^2}
\label{eq:almostH2}
\end{align}
which eventually leads to
\begin{align}
H_{\bm{\omega}/x}\bm{m} = 0.
\label{eq:zero_det}
\end{align}
This confirms that, not only there is a connection among the spectra of the Bethe-Hessian and non-backtracking matrices, but also between their eigenvectors. Note that, by choosing $\omega_{ij} = \xi$ is there if $t$ such that $i,j \in \mathcal{V}_t$ and $\omega_{ij} = h$ otherwise, we precisely recover the definitions of $B_{\xi, h}$ and $H_{\xi/x,h/x}$ as per Equations~(\ref{eq:BHMatrix}, \ref{eq:matrix_B}).

We now further comment how the spectra of $B_{\xi,h}$ and $H_{\xi/y, h/y}$ are related when $y \in \mathbb{R}$ is not an eigenvalue of $B_{\xi, h}$. First recall that, as per Assumption \ref{ass:B}, the large majority of the eigenvalues of $B_{\xi,h}$ are asymptotically bounded by a circle in the complex plane and that only few isolated eigenvalues are larger in modulus with associated eigenvectors representative of the mesoscale structure of $\mathcal{G}$. First consider the case where $y \to \infty$. Then, letting $\tilde{\xi} = \xi/y \to 0$ and $\tilde{h} = h/y \to 0$, by definition (Equation~\eqref{eq:BHMatrix}), it comes that $H_{\tilde{\xi},\tilde{h}} \succ 0$, \emph{i.e.}, all the eigenvalues are positive. %Further note that $y > \rho(B_{\xi,h})$. 
Now, decreasing $y$ to $y = \rho(B_{\xi,h})$, from Equation~\eqref{eq:zero_det}, $H_{\tilde{\xi},\tilde{h}}$ has one eigenvalue equal to zero, which is necessarily the smallest and for all $y > \rho(B_{\xi,h})$, $H_{\tilde{\xi},\tilde{h}}$ is positive definite. This is because if there was a $y > \rho(B_{\xi,h})$ such that ${\rm det}[H_{\tilde{\xi},\tilde{h}}] = 0$, then $y$ would have to be an eigenvalue of $B_{\xi,h}$, which is absurd by construction.

For $y$ lying between the first and the second largest real eigenvalues of $B_{\xi,h}$, no eigenvalue of $H_{\tilde{\xi},\tilde{h}}$ is equal to zero, and the smallest one is negative and isolated. Further decreasing the value of $y$, the smallest (isolated) eigenvalues of $H_{\tilde{\xi},\tilde{h}}$ become progressively negative in correspondence of the largest isolated eigenvalues of $B_{\xi,h}$. 

Formally, this discussion may be summarized as follows.
\begin{property}
	Let $L_{\xi,h}$ be the radius of the bulk of $B_{\xi,h}$ and let $y \geq L_{\xi,h}$. Then, the number of real (isolated) eigenvalues of $B_{\xi,h}$ which are greater (or equal) to $y$ is equal to the number of (isolated) eigenvalues of $H_{\xi/y,h/y}$ which are smaller (or equal) to zero. In particular, for $y = L_{\xi,h}$, the left edge of  bulk spectrum of  $H_{\xi/y,h/y}$ is asymptotically close to $0^+$.
	\label{property}
\end{property}

\begin{figure}[t!]
	\centering
	\includegraphics[width=\columnwidth]{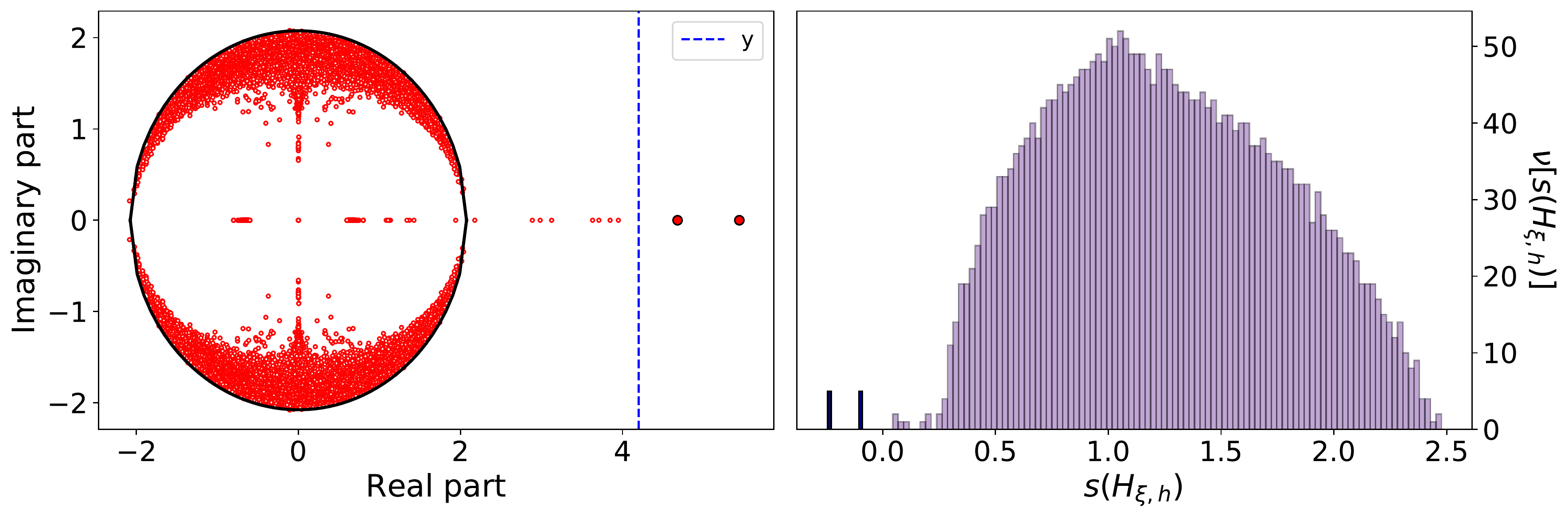}
	\caption{\textbf{Left} : spectrum of the matrix $B_{\xi,h}$ in the complex plane. In blue the considered value of $y$ and in larger size the two eigenvalues of $B_{\xi,h}$ larger than $y$. \textbf{Right} : histogram of $H_{\xi/y,h/y}$ with evidenced the two negative eigenvalues. \textbf{For both simulations}: $n = 1~000$, $T=3$, $k=4$, $c = 6$, $c_{\rm out} = 2$ for all off-diagonal elements of $C$, $\Phi = 1$, $\eta = 0.9$, $\xi = 0.8$, $h = 0.6$ and $y = 4.2$.}
	\label{fig:prop1}
\end{figure}

A pictorial representation of Property~\ref{property} is given in Figure~\ref{fig:prop1}. With this result, we know how to relate the spectrum of $H_{\xi,h}$ to the spectrum of $B_{\xi, h}$ that we study in the next section. 

\subsection{Spectrum of the weighted non-backtracking matrix}
\label{subapp:proof.spectrum}

We now proceed in our agenda by studying the spectrum of $B_{\xi, h}$ under Assumption~\ref{ass:B}. The method we use can be seen as a generalization of \cite{krzakala2013spectral}. By considering the expression of the expected eigenvector, we first determine the position of the eigenvalues belonging to the \emph{informative family} (starting from the largest) and then of the \emph{uninformative family}. Secondly, we analyze the variance of the expression of the expected eigenvector and see under what condition the expectation is meaningful. With this result we finally determine the value of $L_{\xi,h}$ (the radius of the bulk of $B_{\xi,h}$) and summarize our findings in Proposition~\ref{prop:0}.

\subsubsection{The position of the informative eigenvalues}
\label{subsubapp:info}

In this section we determine the position of the informative eigenvalues of $B_{\xi,h}$ with modulus larger than $L_{\xi,h}$. To do so, we first study the largest of them in the limiting case $T\to\infty$, to then extend our findings for finite $T$ to all other eigenvalues.

\textbf{The limiting case of $T\to\infty$}

Consider the graph $\mathcal{G}$ generated according to Definition~\ref{def:G}. Let $\omega_{ij} = \xi$ if there exists $t$ such that $i, j \in\mathcal{V}_t$ and $\omega_{ij} = h$ otherwise, and 
%. We recall the definition of $B_{\xi,h}$ and 
let $\bm{g}^{(r)} \in \mathbb{R}^{|\mathcal{E}|^d}$, for $r \in \mathbb{N}$, be the vector with entry
\begin{align}
%\left(B_{\xi,h}\right)_{(ij),(kl)} &= \mathds{1}_{jk}(1-\mathds{1}_{il})~\omega_{kl} \\
g_{ij}^{(r)} &= \frac{1}{\mu_1^r} \sum_{(wx) ~ :~\underrel{k \neq i}{d(jk,wx) = r}}W_{(jk)\to(wx)}\sigma_x,
\label{eq:guess_vec_B}
\end{align}
where $\{(jk)~:~d(jk,wx)=r\}$ is the set of directed edges $(jk)$ such that the shortest directed non-backtracking path connecting $(jk)$ to $(wx)$ is of length $r$, and where $W_{(jk)\to(wx)}$ is the ``total weight'' of this shortest path defined as the product of each edge weight $\omega_{ij}$, \emph{i.e},
\begin{align}
W_{(jk)\to(wx)}=\omega_{(jk)}\omega_{(k\cdot)}\cdots \omega_{(\cdot w)}\omega_{(wx)}.
\end{align}
The quantity $\sigma_x \in\{\pm1\}$ takes its value according to the label of node $x$.  The value of $\mu_1$ appearing in Equation \eqref{eq:guess_vec_B} will be chosen in order to enforce the vector $\bm{g}^{(r)}$ to be an approximate eigenvector of $B_{\xi,h}$, defined in Equation~\eqref{eq:new_B}. By the definition of $\bm{g}^{(r)}$, recalling the expression of $B_{\xi,h}$ in \eqref{eq:new_B}, we find that
\begin{align}
(B_{\xi,h}\bm{g}^{(r)})_{ij} = \mu_1 g_{ij}^{(r+1)}.
\label{eq:approx_eig}
\end{align}

We now analyze this expression exploiting the tree-like approximation elaborated in Appendix~\ref{app:detectability}. Resuming from this approximation, the expectation of $g_{ij}^{(r)}$ may be written under the following form:
\begin{align}
\mathbb{E}[g_{ij}^{(r)}] = \frac{1}{\mu_1^r}\left(c\Phi\lambda \xi\chi_s^{(r-1)} + \phi_i \eta h \chi_t^{(r-1)}\right)\sigma_j.
\label{eq:expected_g}
\end{align}
Here the first addend is the contribution of the spatial children of $j$ which are on average $c\Phi$ in number, and for each of them the weight of the connecting edge is equal to $\xi$ while the correlation between the labels $\lambda = \mathbb{E}[\sigma_j\sigma_k]$. Each spatial child being at a distance $r-1$ from the target edges -- themselves at a distance $r$ from $(jk)$ -- contributes to the sum through a term which we denoted $\chi_s^{(r-1)} > 0$. Similarly, the second addend is the contribution of the temporal children which are $\phi_i = 2$ in number if $(ij)$ is a spatial edge or $\phi_i = 1$ if $(ij)$ is a temporal edge; their own contribution is denoted $\chi_t^{(r-1)} > 0$. The correlation of the labels of temporal children is equal to $\eta$ and the weight of the edges is equal to $h$. Importantly note that, as a consequence of $\lambda, \xi, \eta, h$ being assumed to be all positive, both $\chi_s^{(r)}$ and $\chi_t^{(r)}$ are positive as well.

By recurrence, the values of $\chi_{s/t}^{(r)}$, which we just defined, then undergo the following relation
\begin{align}
\begin{pmatrix}
\chi_s^{(r)} \\
\chi_t^{(r)}
\end{pmatrix} &=
\begin{pmatrix}
c\Phi\lambda\xi & 2\eta h \\
c\Phi\lambda\xi & \eta h
\end{pmatrix}
\begin{pmatrix}
\chi_s^{(r-1)} \\
\chi_t^{(r-1)}
\end{pmatrix} = 
\begin{pmatrix}
c\Phi\lambda\xi & 2\eta h \\
c\Phi\lambda\xi & \eta h
\end{pmatrix}^r
\begin{pmatrix}
\chi_s^{(0)} \\
\chi_t^{(0)}
\end{pmatrix} \\
&\equiv \left(M_{\infty}(\sqrt{c\Phi\lambda\xi},\sqrt{h\eta})\right)^r    \begin{pmatrix}
\chi_s^{(0)} \\
\chi_t^{(0)}
\end{pmatrix},
\label{eq:relation_chi}
\end{align}
where $M_{\infty}(\cdot,\cdot)$ is the matrix introduced in Equation~\eqref{eq:matrix_M}. For simplicity we will denote it as $M_{\infty}$.
For, say, $r \sim {\rm log}(n)$, $\chi_{s/t}^{(r)} \approx \rho^r(M_{\infty}) v_{s/t}$, where $\bm{v}=(v_s,v_t)$  is the eigenvector associated to the eigenvalue of $M_{\infty}$ of largest amplitude. Equation~\eqref{eq:expected_g} can therefore be further approximated as 
\begin{align}
\mathbb{E}[g_{ij}^{(r)}] = \left(\frac{\rho(M_{\infty})}{\mu_1}\right)^r \left(c\Phi \lambda\xi v_s + \phi_i \eta h v_t\right)\sigma_j + o\left(\frac{\rho(M_{\infty})}{\mu_1}\right)^r.
\end{align}
This expression naturally leads to the choice $\mu_1 = \rho(M_{\infty})$ for which $\mathbb{E}[g_{ij}^{(r)}]$ is independent of $r$, thus turning Equation~\eqref{eq:approx_eig} into an approximate eigenvector equation and $\mu_1$ into a close approximation of one of the real eigenvalues of $B_{\xi,h}$.

\medskip

We now extend this result to the case of finite $T$, and bring further conclusion on all the eigenvalues of $B_{\xi,h}$ belonging to the \emph{informative family}.

\medskip

\textbf{The case of finite $T$}

As we discussed already along Appendix~\ref{app:detectability}, the case of finite $T$ introduces further difficulties due to the time-boundaries $t = 1$ and $t = T$. This being accounted for, when analyzing the contribution of each edge, not only we have to distinguish between spatial and temporal edges, but also to specify the time at which the edge lives. More precisely, suppose that $j \in \mathcal{V}_t$ for $1 \leq t \leq T$. We can rewrite Equation~\eqref{eq:expected_g} as
\begin{align}
\mathbb{E}[g_{ij}^{(r)}] = \frac{1}{\mu_1}\left[c\Phi\lambda\xi\chi_{s,t}^{(r-1)} + (1-\delta_{1,t})\eta h \chi_{b,t}^{(r-1)}+ (1-\delta_{T,t})\eta h\chi_{f,t}^{(r-1)}\right],
\end{align}
where $\chi_{s,t}^{(\cdot)}, \chi_{b,t}^{(\cdot)}, \chi_{f,t}^{(\cdot)}$ are respectively the contributions to the of a spatial, a backwards temporal and a forward temporal child of a node $j \in \mathcal{V}_t$. The relation between all the $\chi$'s directly unfolds from the branching process at finite $T$ that we already discussed in Appendix~\ref{app:detectability}. More precisely, let $\bm{\chi}^{(r)}~=~\{\chi_{b,t}^{(r)}, \chi_{s,t}^{(r)},\chi_{f,t}^{(r)}\}_{t = 1,\dots,T}$, then the following relation holds:
\begin{align}
\bm{\chi}^{(r)} = M_T\left(\sqrt{c\Phi\lambda\xi},\sqrt{\eta h}\right)\bm{\chi}^{(r-1)},
\end{align}
where $M_T(\cdot,\cdot)$ is the matrix defined in Equation~\eqref{eq:transition}. Following the argument we just detailed for $T\to \infty$, we then get that the largest eigenvalue of the \emph{informative family} is asymptotically close to $\mu_1 = \rho\left( M_T\left(\sqrt{c\Phi\lambda\xi},\sqrt{\eta h}\right)\right)$.

\medskip

This analysis also allows us to describe the subsequent eigenvalues $\mu_{i \geq 2}$ belonging to the \emph{informative family} that have a smaller modulus. These modes are metastable configurations of the branching process as in configuration 4 of Figure~\ref{fig:stable}. In these modes, nodes belonging to different communities are still distinguished (hence the reason why these modes are \emph{informative}), but the class identification $\sigma_x$ may be reversed across time. This results in a state in which neighbours are more likely to change label than to keep it, hence they have \emph{negative} label correlation and lead to negative values of $\chi$. This means to relax the constraint $\chi>0$ and thus no longer looking for the leading eigenvalue of $M_T$. From this intuition we argue that the subsequent informative eigenvalues of $B_{\xi,h}$ coincide with the subsequent eigenvalues of $M_T\left(\sqrt{c\Phi\lambda\xi},\sqrt{\eta h}\right)$. 

\medskip

A further important remark should be made on the eigenvalues $\mu_{i \geq 1}$. The matrix $M_T$ is real and non-negative, but it is not symmetric. Consequently, the leading eigenvalue, $\mu_1$ will certainly be real (due to Perron-Frobenius theorem), while the subsequent eigenvalues are potentially complex. Although we cannot offer a clear interpretation for the complex nature of some of these isolated eigenvalues, our study is experimentally verified to hold also in this case as shown in Figure~\ref{fig:spectrumBT}.

\medskip

We now proceed extending our arguments to the \emph{uninformative family} of isolated eigenvalues of $B_{\xi,h}$.

\subsubsection{The position of the uninformative isolated eigenvalues}

As in the static case, not all stable configurations of the branching process of Appendix~\ref{app:detectability} are informative. In particular, two nodes of $\mathcal{G}$ might be considered to belong to the same community only because they live at the same time. Based on the technique detailed in Section~\ref{subsubapp:info}, we now describe the position of the eigenvalues forming the \emph{uninformative family}. Although these eigenvalues are not informative, the awareness of their presence is crucial if one has to avoid to mistakenly use one of these for community reconstruction. 

\medskip

We proceed again by studying the largest of these eigenvalues (which is also the largest eigenvalue of $B_{\xi,h}$), to then extended the result to all the others. Let us denote $\{\gamma_i\}_{i = 1,\dots,T}$ this second set of (trivial and non-informative) eigenvalues. The approximate Perron-Frobenius eigenvector $\bm{b} \in\mathbb{R}^{2|\mathcal{E}|}$ can be written as
\begin{align}
b_{ij}^{(r)} &= \frac{1}{\gamma_1^r} \sum_{(wx) ~ :~\underrel{k \neq i}{d(jk,wx) = r}}W_{(jk)\to(wx)}.
\label{eq:expected_b}
\end{align}
According to this expression, we set $\sigma_x = 1$ for all nodes and thus the correlation between $\sigma_x$ and $\sigma_y$ is always unitary.
Following the argument developed to determine the value of $\mu_1$, we then obtain
\begin{align}
\gamma_1 = \rho\left(M_T\left(\sqrt{c\Phi\xi},\sqrt{h}\right)\right).
\end{align}
As in Section~\ref{subsubapp:info}, this eigenvalue is necessarily real and the subsequent eigenvalues of the 	\emph{uninformative family} are given by the subsequent eigenvalues of $M_T\left(\sqrt{c\Phi\xi},\sqrt{h}\right)$ and can be complex. Note importantly that the ordering of $\{\mu_i\}_{i\geq 1}$ and $\{\gamma_i\}_{i\geq 1}$ is not \emph{a priori} well defined.

\medskip

So far we determined the position of the isolated eigenvalues under the assumption that the expectation of the approximate eigenvectors are significant. In order to know when this analysis holds, we have to study the variance of the entries of the approximate eigenvectors and see under what conditions it vanishes. This analysis will also allow us to determine the value of the radius of the bulk of $B_{\xi,h}$. 

\subsubsection{The bulk eigenvalues of $B_{\xi,h}$}

To begin with, we investigate under which conditions the approximate eigenvector Equations~(\ref{eq:expected_g}, \ref{eq:expected_b}) hold. We then proceed with a study of the variance of $g_{ij}^{(r)}$ (and $b_{ij}^{(r)}$). When the variance vanishes, the eigenvector is well approximated by its expectation and we conjecture it is isolated. On the contrary, when the variance diverges it is because it gets asymptotically close to the bulk of uninformative eigenvalues and is no longer isolated.

Let us first consider the eigenvector attached to $\mu_1$:
\begin{align}
\label{eq:E[g2]}
\mathbb{E}\left[\left(g_{ij}^{(r)}\right)^2\right] = \frac{1}{\mu_1^{2r}}\sum_{(wx)~:\underrel{k \neq i}{~d(jk,wx) = r}} \left(W_{(jk)\to(wx)}^2 + \sum_{(vy)~:~\underrel{(vy)\neq(wx), k\neq i}{d(jk,vy) = r}}\sigma_x\sigma_y W_{(jk)\to(wx)}W_{(jk)\to(vy)}\right).
\end{align}
The first addend of \eqref{eq:E[g2]} can be evaluated as previously done in Equation~\eqref{eq:relation_chi}, getting
\begin{align}
\mathbb{E}\left[\frac{1}{\mu_1^{2r}}\sum_{(wx)~:\underrel{l \neq i}{~d(jl,wx) = r}} W_{(jl)\to(wx)}^2 \right]= O\left(\frac{\rho^r\left(M_T(\sqrt{c\Phi\xi^2},h)\right)}{\mu_1^{2r}}\right).
\end{align}
If $\mu_1^2 < \rho\left(M_T(\sqrt{c\Phi\xi^2},h)\right)$, this addend of \eqref{eq:E[g2]} diverges, and so does the variance of $g_{ij}^{(r)}$: in this case, $\bm{g}^{(r)}$ cannot be an approximate eigenvector of $B_{\xi,h}$.

Consider next the second addend of Equation~\eqref{eq:E[g2]}:
\begin{align}
&\mathbb{E}\left[\frac{1}{\mu_1^{2r}}\sum_{(wx)~:\underrel{l \neq i}{~d(jk,wx) = r}} \sum_{(vy)~:~\underrel{(vy)\neq(wx), k\neq i}{d(jk,vy) = r}}\sigma_x\sigma_y W_{(jk)\to(wx)}W_{(jk)\to(vy)}\right] \nonumber\\
&= \frac{1}{\mu_1^{2r}}\sum_{(wx)~:\underrel{l \neq i}{~d(jk,wx) = r}} \sum_{(vy)~:~\underrel{(vy)\neq(wx), k\neq i}{d(jk,vy) = r}} \mathbb{E}[\sigma_j\sigma_xW_{(jk)\to(wx)}\cdot\sigma_j\sigma_yW_{(jk)\to(vy)}] \nonumber \\
&\approx \frac{1}{\mu_1^{2r}}\sum_{(wx)~:\underrel{l \neq i}{~d(jk,wx) = r}} \mathbb{E}[\sigma_j\sigma_xW_{(jk)\to(wx)}]\sum_{(vy)~:~\underrel{(vy)\neq (wx), k\neq i}{d(jk,vy) = r}} \mathbb{E}[\sigma_j\sigma_yW_{(jk)\to(vy)}] \nonumber\\
&\approx \frac{1}{\mu_1^{2r}}\sum_{(wx)~:\underrel{l \neq i}{~d(jk,wx) = r}} \mathbb{E}[\sigma_j\sigma_xW_{(jk)\to(wx)}]\sum_{(vy)~:~\underrel{k\neq i}{d(jk,vy) = r}} \mathbb{E}[\sigma_j\sigma_yW_{(jk)\to(vy)}] \nonumber \\
&= \mathbb{E}^2\left[g_{ij}^{(r)}\right],
\end{align}
where we exploited the fact that the paths $(jk\to wl)$ and $(jk\to vl)$ are asymptotically independent and that the number of paths leading to nodes a distance $r$ from $(jk)$ is exponentially large in $r$, unlike the number of paths leading to $(vy)$ from $(jk)$. We thus obtain that the variance $\mathbb{V}[g_{ij}^{(r)}]$ of $g_{ij}^{(r)}$ grows as
\begin{align}
\mathbb{V}\left[g_{ij}^{(r)}\right] = O\left(\frac{\rho^r\left(M_T(\sqrt{c\Phi\xi^2},h)\right)}{\mu_1^{2r}}\right).
\label{eq:variance_g}
\end{align}
As a consequence, the variance of $g_{ij}^{(r)}$ vanishes if and only if $\mu_1 > \sqrt{\rho\left(M_T(\sqrt{c\Phi\xi^2},h)\right)}$. 

\medskip

Considering now the problem of evaluating the variance for all the $\{\mu_i\}_{i \geq 1}$ and $\{\gamma_i\}_{i\geq 1}$, note that, the variance is only determined by the first addend of Equation~\eqref{eq:E[g2]}. This term does not depend on the configuration $\bm{\sigma}$ and is, therefore, the same for \emph{all} the isolated eigenvectors. Consequently, for all the isolated eigenvectors, the variance vanishes if the corresponding eigenvalue is greater than $L_{\xi,h} = \sqrt{\rho\left(M_T(\sqrt{c\Phi\xi^2},h)\right)}$, which is precisely the radius of the bulk of $B_{\xi,h}$, since an informative eigenvalue-eigenvector pair $(\mu_i,\bm{g}_i)$, (resp. $(\gamma_i, \bm{b}_i)$), for $B_{\xi,h}$ can only exist provided that $\mu_i$ (resp. $\gamma_i$) is greater than $L_{\xi,h}$.

\begin{figure}[t!]
	\centering
	\includegraphics[width = \columnwidth]{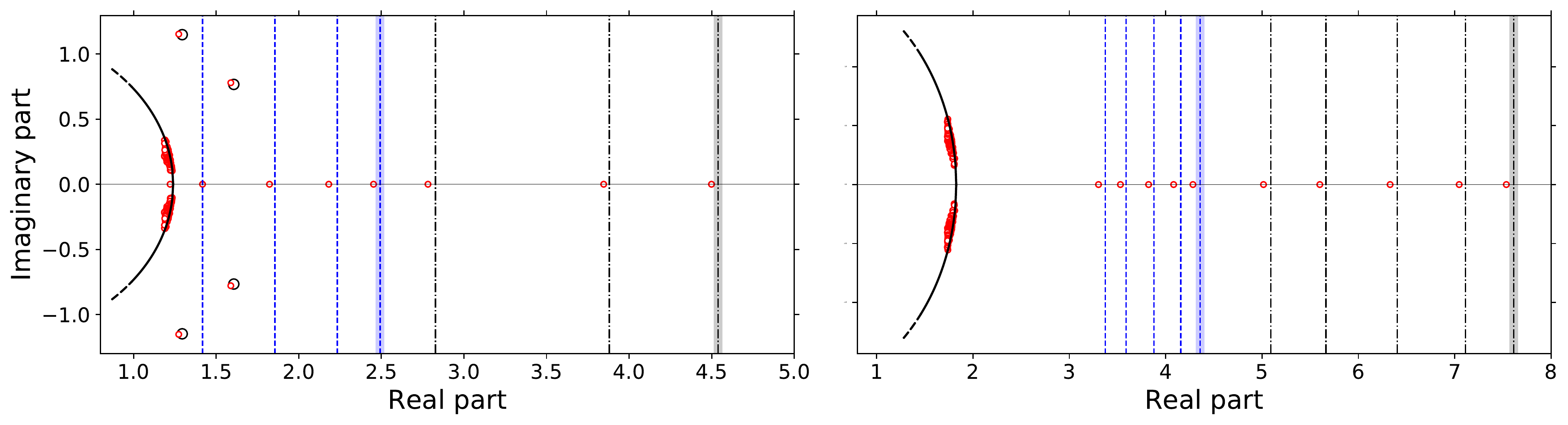}
	\caption{The 150 eigenvalues of $B_{\xi,h}$ with largest real part, for $n = 10\,000$, $T = 5$, $\eta = 0.4$, $c = 10$, $c_{\rm out} = 4$, $\Phi = 1.64$. \textbf{Left} $\xi = 0.2$, $h = 0.9$. \textbf{Right} $\xi = 0.4$, $h = 0.7$. The blue dashed lines are the theoretical positions of the eigenvalues forming the \emph{informative family}, while the black dashed-dotted lines indicate the \emph{uninformative family}. The thickest blue and black lines are $\mu_1$ and $\gamma_1$, respectively. The imaginary eigenvalues are represented with a circle in the complex plane. The solid black line is a part of a circle of radius $L_{\xi,h}$.}
	\label{fig:spectrumBT}
\end{figure}

The results of this section may be summarized as follows.
\begin{prop}
	Letting $\mathcal{G}$ be a graph generated as per Definition~\ref{def:G}, in the $n\to \infty$ limit, the complex eigenvalues forming the bulk of $B_{\xi,h}$ are bounded by a disk in the complex plane of radius $L_{\xi,h} = \sqrt{\rho(M_T(\sqrt{c\Phi\xi^2},h))}$, for $M_T(\cdot,\cdot)$ defined in Equation~\eqref{eq:transition}. All the eigenvalues of $B_{\xi,h}$ of magnitude larger than $L_{\xi,h}$ are isolated and are asymptotically close to one of the eigenvalues of either $M_T(\sqrt{c\Phi\xi\lambda},\sqrt{\eta h})$ (in which case they correspond to non-trivial modes) or $M_T(\sqrt{c\Phi\xi},\sqrt{h})$ (in which case they correspond to trivial modes).
	\label{prop:0}
\end{prop}

Figure \ref{fig:spectrumBT} confirms numerically Proposition~\ref{prop:0} for two choices of values of $(\xi,h)$, one in which all the isolated eigenvalues are real and one in which there are complex isolated eigenvalues. We choose to compute only the $150$ eigenvalues with largest real part to keep a reasonable computational time, while having a large value of $n$.

Based on these results, we now proceed giving the supporting arguments of Proposition~\ref{prop:2}.

\subsection{Supporting arguments for Proposition \ref{prop:2}}
\label{subapp:proof.prop2}

This section provides the final theoretical support to Proposition~\ref{prop:2} at the core of the article, being at the root of our proposed dynamic clustering algorithm. To this end, we need to show how the bulk spectrum of $B_{\xi,h}$ relates to the bulk spectrum of $H_{\xi,h}$ for the values of $(\xi,h)$ proposed in Proposition~\ref{prop:2}, \emph{i.e.,} $\xi=\lambda_d=\frac{\alpha_c(T,\eta)}{\sqrt{c\Phi}}$ and $h=\eta$.

\medskip

Exploiting the result of Proposition~\ref{prop:0}, the matrix $B_{\lambda_d,\eta}$ has an eigenvector correlated to the class labels equal to $\mu_1 = \rho(M_T(\sqrt{c\Phi\lambda\lambda_d}, \eta))$.
First note that, by definition, $\sqrt{c\Phi\lambda_d^2} = \alpha_c(T,\eta)$, while $\sqrt{c\Phi\lambda^2} = \alpha$. For $\alpha > \alpha_c(T,\eta)$, then $\lambda > \lambda_d$, and, consequently $\sqrt{c\Phi\lambda\lambda_d}>\alpha_c(T,\eta)$. From this last equation and the definition of $\alpha_c(T,\eta)$ provided in Section~\ref{subsec:model.detectability}, we conclude that $\mu_1>1$.

From Proposition~\ref{prop:0}, we further have that the radius of the bulk spectrum of $B_{\lambda_d,\eta}$ is equal to $L_{\lambda_d,\eta} = 1$. As such, the informative eigenvalue $\mu_1$ of $B_{\lambda_d,\eta}$ exists as soon as $\alpha > \alpha_c(T,\eta)$.

From Property~\ref{property}, the smallest eigenvalue of the bulk (\emph{i.e.}, its left-edge) of $H_{\lambda_d,\eta}$ is asymptotically close to zero and all the eigenvectors associated to the negative eigenvalues are correlated to the mesoscale structure of $\mathcal{G}$, thereby entailing the validity and optimal performance down to the detectability threshold of our proposed Algorithm~\ref{alg:0}.

\medskip

In Figure~\ref{fig:spec} (subplots 2 and 4) we provide numerical support to Proposition~\ref{prop:2}, showing the spectra of $B_{\lambda_d,\eta}$ and $H_{\lambda_d,\eta}$.

\subsection{Analysis of the spectrum of $B_{\lambda,\eta}$}
\label{subapp:proof.prop1}

In the previous sections we studied the spectrum of $B_{\xi,h}$ for generic $(\xi,h)$. We now focus on the particular choice $(\xi = \lambda, h = \eta)$ that leads to $B_{\lambda,\eta}$, sharing the same eigenvalues of the dynamical non-backtracking of \cite{ghasemian2016detectability}. First we show that this matrix has an informative isolated eigenvalue (not necessarily the second largest) for all $\alpha > \alpha_c(T,\eta)$. We then show that the matrix $H_{\lambda,\eta}$ shares the same property. We further comment that, however, the choice $(\xi = \lambda, \eta = h)$ is impractical from an algorithmic standpoint.
\newpage

\textbf{Community detectability with $B_{\lambda,\eta}$}

The fact that the matrix $B_{\lambda,\eta}$ can be used for community reconstruction is a straightforward consequence of Proposition~\ref{prop:0}. In fact, letting $\xi = \lambda$ and $h = \eta$, we obtain that the leading informative eigenvalue is equal to $\mu_1 = \rho(M_T\left(\alpha,\eta)\right)$, while the radius of the bulk is equal to $L_{\lambda,\eta} = \sqrt{\rho(M_T\left(\alpha,\eta)\right)} = \sqrt{\mu_1}$. By definition, if $\alpha > \alpha_c(T,\eta)$, then $\mu_1 > 1$, therefore $\mu_1 > L_{\lambda,\eta}$. So for all $\alpha > \alpha_c(T,\eta)$, $\mu_1$ is an isolated eigenvalue in the spectrum of $B_{\lambda,\eta}$, but it does not correspond, in general, to the second largest eigenvalue.

	\begin{figure}[t!]
	\centering
	\includegraphics[width = \columnwidth]{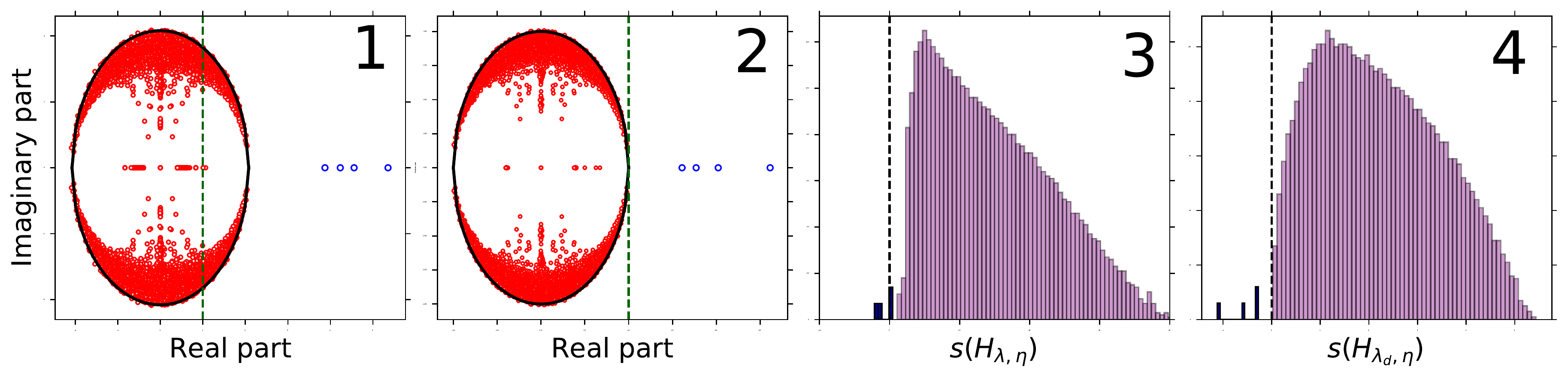}
	\caption{\textbf{Sub-figures 1, 2}: spectrum of $B_{\xi,\eta}$ for $\xi = \lambda$ and $\xi = \lambda_d$, respectively. The green dashed line is the position of $1$, while the black circle is of radius $L_{\xi,\eta}$. \textbf{Sub-figures 3, 4}: histogram of $H_{\xi,\eta}$ for $\xi = \lambda$ and $\xi = \lambda_d$, respectively. The black dashed line indicates the position of $0$. For all simulations, $T = 2$, $\eta = 0.4$, $c = 6$, $c_{\rm out} = 1$, $\Phi = 1$, $n = 2~000$.}
	\label{fig:spec}
\end{figure}

We now proceed our discussion studying the matrix $H_{\lambda,\eta}$.

\textbf{Community detectability with $H_{\lambda,\eta}$}

In order to fully grasp the properties of the matrix $H_{\lambda,\eta}$, one has to consider its relation with $B_{\lambda,\eta}$ and the \emph{belief propagation} (BP) equations. Specifically this allows us to show that the most informative eigenvalue of $B_{\lambda,\eta}$ is $1< L_{\lambda,\eta}$ and lies isolated inside the bulk. Consequently, as per Section~\ref{subapp:BH_B}, the most informative eigenvalue of $H_{\lambda,\eta}$ is equal to zero.

We first establish that $B_{\lambda,\eta}$ naturally comes into play by linearizing BP equations: these consist of a set of fixed-point equations defining ``messages'' $m_{ij}$ exchanged between the nodes $i$ and $j$, and ultimately providing an asymptotically optimal community clustering algorithm. Specifically, from the expression of the whole set of messages $m_{ij}$, one can estimate the marginal probability distribution of the label of each node. To this end, first define
\begin{align}
H = \begin{pmatrix}
\frac{1+\eta}{2} & \frac{1-\eta}{2} \\
\frac{1+\eta}{2} & \frac{1-\eta}{2}
\end{pmatrix}, \quad 
C = \begin{pmatrix}
c_{\rm in} & c_{\rm out} \\
c_{\rm out} & c_{\rm in}
\end{pmatrix}.
\end{align}
Letting $a,b \in \{\pm 1\}$, the BP equations take the form \cite[Equations 5,6,8]{ghasemian2016detectability}
\begin{align}
m_{j_t,i_t}(a) &= \frac{e^{-h_t(a)}}{Z_{j_t,i_t}}\left(\sum_b H_{ab}~m_{i_t,i_{t+1}}(b) \right)\left(\sum_b H_{ab}~m_{i_t,i_{t-1}}(b) \right)\prod_{l_t \in \partial i_t \setminus j_t} \sum_b C_{ab}m_{i_t,l_t}(b) \nonumber\\
m_{i_{t+1},i_t}(a) &= \frac{e^{-h_t(a)}}{Z_{i_{t+1},i_t}}\left(\sum_b H_{ab}~m_{i_t,i_{t-1}}(b) \right)\prod_{l_t \in \partial i_t} \sum_b C_{ab}m_{i_t,l_t}(b)
\end{align}
where
\begin{align}
h_t(a) = \frac{1}{n}\sum_{j \in\mathcal{V}_t}\sum_b C_{ab}m_{i_t,j_t}(b).
\end{align}
The above messages can be expanded around the so-called \emph{trivial} fixed point\footnote{In this fixed point the messages are independent of the class labels, hence it is called \emph{trivial}. From a simple substitution one can indeed verify that it is a fixed point.} $m_{j_t,i_t}(\pm 1) = 1/2 \pm \epsilon_{i_t,j_t}$, $m_{i_t,i_{t\pm 1}}(\pm 1) = 1/2 \pm \epsilon_{i_t,i_{t\pm 1}}$, yielding
\begin{align}
\epsilon_{j_t,i_t} &= \eta(\epsilon_{i_t,i_{t-1}}+\epsilon_{i_t,i_{t+1}})+\lambda \sum_{\ell_t \in \partial i_t \setminus j_t} \epsilon_{i_t,\ell_t} \\
\epsilon_{i_{t+1},i_t} &= \eta\epsilon_{i_t,i_{t-1}}+\lambda \sum_{\ell_t \in \partial i_t} \epsilon_{i_t,\ell_t}.
\end{align}
These equations can be rewritten in synthetic form introducing the weighted non-backtracking matrix 
\begin{align}
B_{\lambda,\eta}\bm{\epsilon} = \bm{\epsilon}.
\label{eq:informative_B}
\end{align}
In agreement with our empirical observations, we predict that the matrix $B_{\lambda,\eta}$ has an eigenvalue asymptotically close to one, so that, as a consequence of the property discussed in Appendix \ref{subapp:BH_B}, $H_{\lambda,\eta}$ has an eigenvalue asymptotically close to zero. The corresponding eigenvector of $B_{\lambda,\eta}$ represents the deviation from the trivial fixed point and is naturally connected to the community structure. The presence (and importance) of this isolated eigenvalue has been already observed and studied in the static regime \cite{dall2019revisiting,coste2019eigenvalues} and is visually depicted in Figure~\ref{fig:spec} (subplots 1 and 3).

We finally argue that this eigenvalue of $B_{\lambda,\eta}$ exists and is isolated as soon as $\alpha > \alpha_c(T,\eta)$. Indeed, the eigenvalue equal to one lies \emph{isolated inside} the bulk of $B_{\lambda,\eta}$, the radius of the bulk spectrum of $B_{\lambda,\eta}$ being $L_{\lambda,\eta} = \sqrt{\rho(M_T(\alpha,\eta))}$. There further exists another informative eigenvalue which is equal to $\mu_1 = \rho(M_T(\alpha,\eta))$.  The eigenvalue equal to $1$ remains isolated inside the bulk for all $\alpha > \alpha_c(T,\eta)$ and meets the outer-bulk isolated eigenvalue, $\mu_1$, right at the edge of the bulk when $\alpha = \alpha_c(T,\eta)$ (\emph{i.e.}, at the precise detection threshold). Below the transition threshold, when $\alpha<\alpha_c(T,\eta)$, the two eigenvalues then become complex conjugate.

This result can be summarized in the form of the following proposition.

\begin{prop}
	\label{prop:1}
	Let $\mathcal{G}$ be a graph generated as per Definition \ref{def:G}. As $n \to \infty$, the complex eigenvalues forming the bulk of of the non-symmetric matrix $B_{\lambda,\eta}$ are asymptotically bounded by a circle in the complex plane of radius $L_{\lambda,\eta} = \sqrt{\rho(M_T(\alpha,\eta))}$, with $\alpha = \sqrt{c\Phi\lambda^2}$ and $M_T(\alpha,\eta)$ defined in \eqref{eq:matrix_M}.
	
	Besides, if $\alpha > \alpha_c(T,\eta)$, then $1<L_{\lambda,\eta}$, $1$ is an isolated eigenvalue of $B_{\lambda,\eta}$ and $0$ is an isolated eigenvalue of $H_{\lambda,\eta}$, and the corresponding eigenvectors for both matrices are correlated to the vector of community labels.
\end{prop}

Proposition~\ref{prop:1} states that one informative eigenvector of $H_{\lambda,\eta}$ (the one corresponding to the mode 2 of Figure~\ref{fig:stable}) is associated to the zero eigenvalue, but nothing is said on its relative position in the spectrum of $H_{\lambda,\eta}$. This is a practical issue: indeed, as $\lambda$ is also a priori unknown, one cannot simply browse over values of $\lambda$ in search for an isolated zero eigenvalue of $H_{\lambda,\eta}$, which may correspond to a non-informative mode. 
\medskip

The numerical support of Proposition~\ref{prop:1} (subplots 1 and 3) is provided by Figure~\ref{fig:spec}.

\section{Dependence of the realizations of $A^{(t)}$ by adding edge persistence}
\label{app:memory}

This section provides hints to generalize the main results of the article to networks with persistence not only in the labels, but also in the links that can be maintained across successive (therefore non longer independent) realizations of the graph. Link persistence has a deleterious effect on community detection because it introduces \emph{lagged inference} \cite{barucca2018disentangling,barucca2017detectability}, \emph{i.e.}, the reconstruction at time $t$ accounts for the realization of the network at earlier than present time. Specifically, the following generative model is now assumed:
\begin{align}
A_{ij}^{(t+1)} = \begin{cases}
A_{ij}^{(t)} \quad {\rm w.p.}~\tau \\
\Delta_{ij}^{(t)} \quad {\rm w.p.}~(1-\tau)
\end{cases} \quad {\rm where} \quad
\Delta_{ij}^{(t)} = 
\begin{cases}
1\quad {\rm w.p.}~\theta_{i}\theta_{j}\frac{C_{\ell_{i_t},\ell_{j_t}}}{n} \\
0 \quad {\rm otherwise}.
\end{cases}
\label{eq:markov_memory}
\end{align}
The scenario covered in Section~\ref{sec:model} of the main article allows one to infer the community structure from $\{\Delta^{(t)}\}_{t = 1,\dots,T}$ but we only observe its "spoiled" version $\{A^{(t)}\}_{t = 1,\dots, T}$. In order to overcome this limitation, we introduce the following matrix:
\begin{align}
\tilde{A}_{ij}^{(t+1)} =
\begin{cases}
1 \quad {\rm if} \quad A_{ij}^{(t+1)} = 1 \quad {\rm and} \quad A_{ij}^{(t)} = 0 \\
0 \quad {\rm else}.
\end{cases}
\end{align}
In other words, if the same link is repeated at two successive time steps, it is deleted, because, if it was repeated, with high probability it must have been copied (recall that the probability of a link to spontaneously appear in our sparse regime is of order $O(1/n)$). Given the sparsity of $\Delta$, the matrices $\tilde{A}_{ij}^{(t+1)}$ and $\tilde{A}_{ij}^{(t)}$ are asymptotically independent and we thus recover the framework considered in Section~\ref{sec:model} of the main article, when using $\tilde{A}^{(t)}$ (instead of $A^{(t)}$), provided that the detectability conditions on $\tilde{A}^{(t)}$ are met. 

Let us investigate this detectability aspect. Starting from
\begin{align}
\mathbb{P}(\tilde{A}_{ij}^{(t+1)} = 1) = \mathbb{P}(A_{ij}^{(t+1)} = 1|A_{ij}^{(t)} = 0)\big(1-\mathbb{P}(A_{ij}^{(t)} = 1)\big)
\label{eq:P_A_tilde}
\end{align}
we compute the value of $\mathbb{P}(A_{ij}^{(t+1)} = 1)$ recursively:
\begin{align}
\mathbb{P}(A_{ij}^{(t+1)} = 1) = \tau\mathbb{P}(A_{ij}^{(t)} = 1) + (1-\tau)~\mathbb{P}(\Delta_{ij}^{(t+1)} = 1)
\label{eq:result_recurive1}
\end{align}
and thus, from time $t=1$,
\begin{align}
\mathbb{P}(A_{ij}^{(t+1)} = 1) = \sum_{m = 1}^t \mathbb{P}\left(\Delta_{ij}^{(t+1-m)} = 1\right) \tau^m  -  \sum_{m = 1}^{t-1} \mathbb{P}\left(\Delta_{ij}^{(t+1-m)} = 1\right) \tau^{m+1}= O_n\left(\frac{1}{n}\right).
\label{eq:result_recurive}
\end{align}
Hence, injecting Equation \ref{eq:result_recurive} into Equation \ref{eq:P_A_tilde}, we obtain
\begin{align}
\mathbb{P}(\tilde{A}_{ij}^{(t+1)} = 1) = \mathbb{P}(A_{ij}^{(t+1)} = 1 |A_{ij}^{(t)} = 0)(1+o_n(1)) = (1-\tau)\theta_{i}\theta_{j}\frac{C_{\ell_{i_t},\ell_{j_t}}}{n} + o_n(1).
\end{align}
The generative model of $\tilde{A}_{ij}^{(t)}$ thus asymptotically follows a DC-SBM in which the entries of $C$ are multiplied times $(1-\tau)$.

\begin{figure}[t!]
	\centering
	\includegraphics[width=\columnwidth]{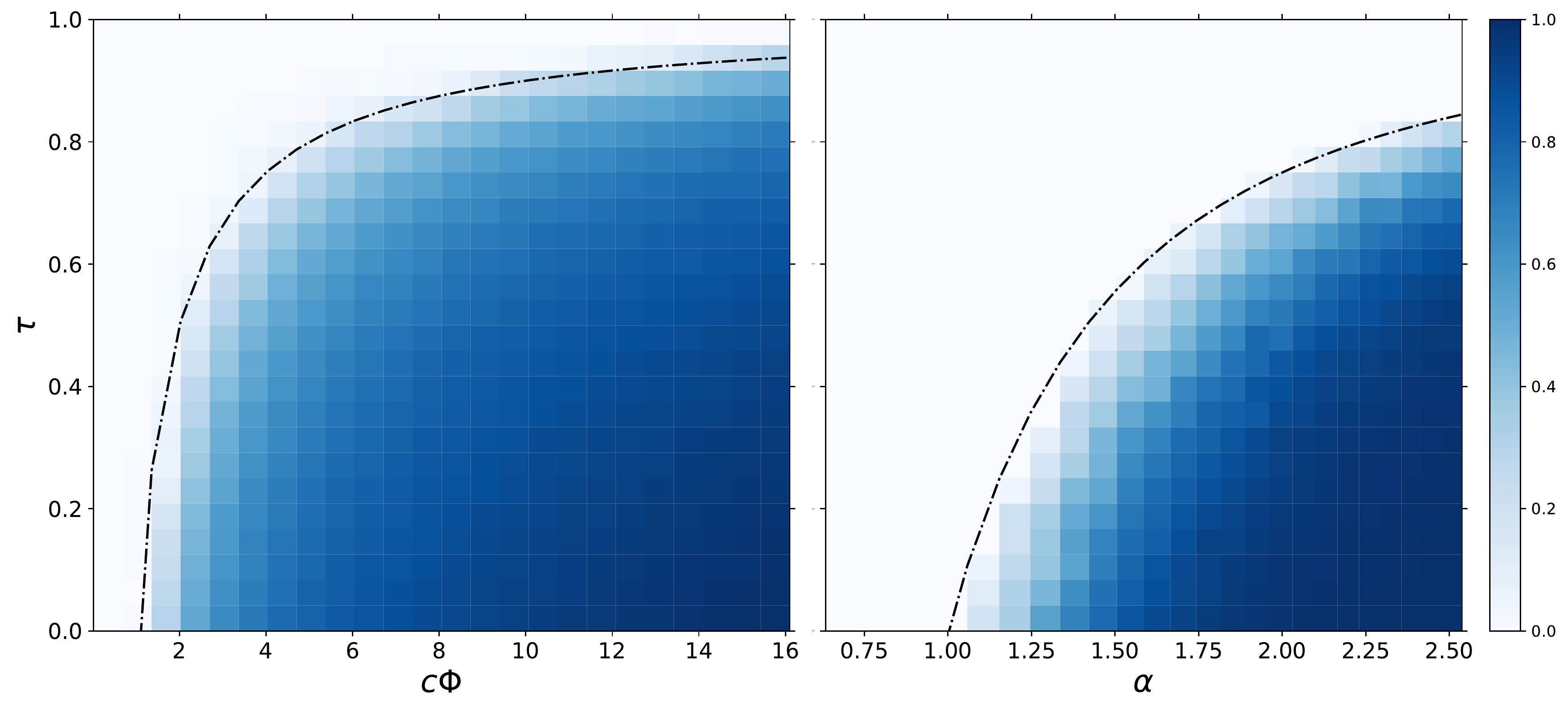}
	\caption{\textbf{Left}: Size of the biggest connected component divided by $n$ as a function of $c$ and $\tau$. The black line indicates the theoretical percolation threshold on the matrix $\tilde{A}$, Equation \eqref{eq:giant_memory}. \textbf{Right} Plot of ${\rm th}(3.5(\hat{\alpha}-1))$ (see text) as a function of $\alpha$ and $\tau$. The black line indicates the static detectability threshold on the matrix $\tilde{A}$, Equation \eqref{eq:detectability_memory}. \textbf{For both figures}: $T = 15$, $n = 5~000$, $k = 2$, $\Phi = 1.65$, $\eta = 0.8$. Averages are taken over three samples.}
	\label{fig:memory}
\end{figure}

To test our theoretical analysis, we evaluate numerically the percolation threshold and the detectability threshold on the matrix $\tilde{A}^{(T)}$. More specifically, the percolation threshold defines the condition under which the graph corresponding to $\tilde{A}^{(T)}$ has a giant component. For the DC-SBM (which generates $\Delta^{(T)}$), this condition is met whenever $c\Phi > 1$ \cite{dall2020unified}. The generative model of $\tilde{A}^{(T)}$ is asymptotically a DC-SBM in which all entries of the matrix $C$ are multiplied times a factor $(1-\tau)$. The percolation threshold hence becomes
\begin{align}
(1-\tau)c\Phi > 1
\label{eq:giant_memory}
\end{align}
In the left plot of Figure~\ref{fig:memory} we generated, for different values of $c\Phi$ and different values of $\tau$, a sequence of $T=15$ snapshots according to Equation~\eqref{eq:markov_memory} and plotted in color code the size of the giant component of $\mathcal{G}_T$, divided by the size of the graph. The dash-dotted black line indicates the position of the percolation threshold that evidences a good agreement between the theoretical prediction and the numerical experiment.

\medskip

Concerning the detectability threshold, instead, the updated (static) detectability threshold here reads
\begin{align}
\alpha > \frac{1}{\sqrt{1-\tau}}.
\label{eq:detectability_memory}
\end{align}
In order to estimate $\alpha$ we compute $\hat{\alpha}$

\begin{align}
\hat{\alpha} &= \frac{\hat{c}_{\rm in}-\hat{c}}{\sqrt{\hat{c}}}\sqrt{\hat{\Phi}},
\end{align}
where
\begin{align}
\hat{c} = \frac{1}{n}\sum_{i,j \in \mathcal{V}_T} \tilde{A}_{ij}^{(T)}; \quad \hat{c}_{\rm in} &= \frac{2}{n} \sum_{i,j \in \mathcal{V}_T : \ell_i = \ell_j} \tilde{A}_{ij}^{(T)}; \quad \hat{\Phi} = \frac{1}{n\hat{c}^2}\sum_{i \in \mathcal{V}_T}\left(\sum_{j \in\mathcal{V}_T} \tilde{A}_{ij}^{(T)}\right)^2
\end{align}
With a similar procedure as the one described to evaluate numerically the percolation threshold, in the right subplot of Figure~\ref{fig:memory}, we display in color code the value of ${\rm th}(3.5(\hat{\alpha}-1))$, saturating the negative values to zero. When $\hat{\alpha} > 1$ the plotted function is between zero and one and we are above the transition. On the opposite, when $\hat{\alpha} < 1$ we are below the transition. The black dash-dotted line confirms the theoretical prediction of the detectability threshold, confirming also in this case our theoretical results.

Concluding, to get rid of the lag effect introduced by the persistence in the edges, one needs to remove at each time step the edges that are repeated. The positions of the information-theoretic transitions are asymptotically the same as those of a D-DCSBM model in which the entries of the matrix $C$ are re-scaled by a factor $1-\tau$, the proportion of edges that do not get copied.

\section{Performance comparison}
\label{app:simulation}

\begin{figure}
    \centering
    \includegraphics[width = \columnwidth]{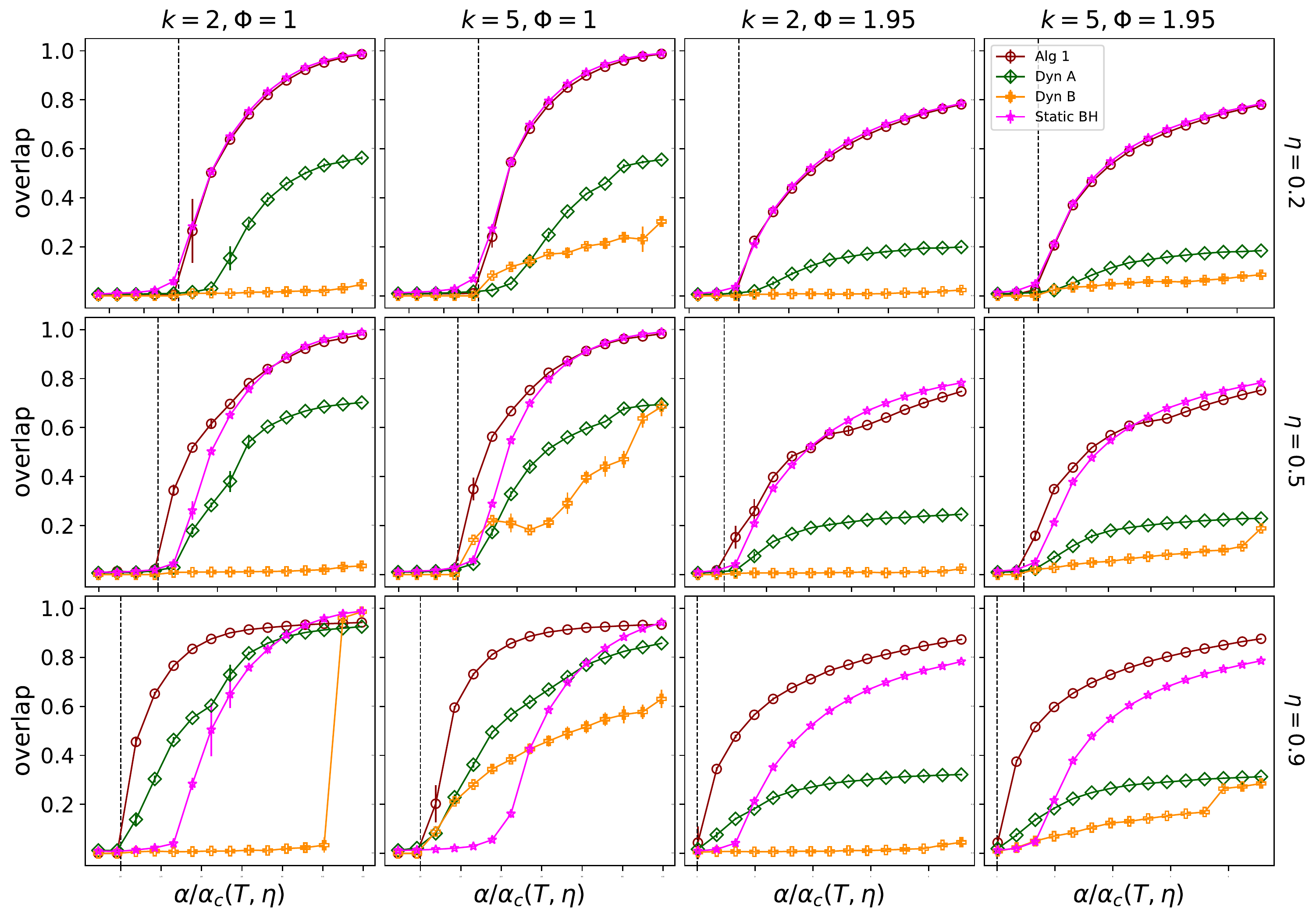}
    \caption{Overlap comparison of Algorithm~\ref{alg:0} (Alg 1), the dynamic adjacency matrix of \cite{keriven2020sparse} (Dyn A), the dynamic non-backtracking of \cite{ghasemian2016detectability} (Dyn B) and the static Bethe-Hessian of \cite{dall2019revisiting} (Static BH). The title of each row ant column indicates the values of $\eta, k, \Phi$ considered. For $\Phi \neq 1$ a power law degree distribution is adopted. The value of $\alpha$ is defined as $\alpha = \sqrt{c\Phi\lambda^2}$, where $\lambda = (c_{\rm in} - c_{\rm out})/(kc)$. The vertical line indicates the position of $\alpha/\alpha_c(T,\eta) = 1$. \textbf{For all simulations}: $c = 6$, $c_{\rm out} = 0.5 \to 5$, $n = 25~000$, $T = 4$. Averages are taken over $10$ samples.}
    \label{fig:big_comparison}
\end{figure}

This section compares numerically the performance of Algorithm~\ref{alg:0} against the main spectral methods commented along the paper. In Figure~\ref{fig:big_comparison} the algorithms are tested for a different number of classes, value of $\eta$ and degree distribution. For $k > 2$ a symmetric setting with classes of equal size and $C_{ab} = c_{\rm out}$ for all $a \neq b$ is considered, so that the spectral algorithm of \cite{ghasemian2016detectability} is still well defined. Figure~\ref{fig:big_comparison} indeed confirms that Algorithm~\ref{alg:0} (i) benefits from high label persistence $\eta$; (ii) systematically outperforms the two considered competing dynamical sparse spectral algorithms \cite{keriven2020sparse}, \cite{ghasemian2016detectability}; (iii) is capable of handling an arbitrary degree distribution.

To compare the performance of Algorithm~\ref{alg:0} and the static Bethe-Hessian of \cite{dall2019revisiting}, the case of small and large values of $\alpha$ should be treated separately. Close to the transition, Algorithm~\ref{alg:0} improves over the static Bethe-Hessian and this gets more evident as $\eta$ increases: the joint solution of the problem at all times allows to improve the clustering performance in the hard detection regime. For large values of $\alpha$, instead, there seems to exist $\alpha^*(\eta)$  beyond which regularity only marginally improves the detection performance and Algorithm~\ref{alg:0} performs equally (or slightly worse) than the static algorithm of \cite{dall2019revisiting}. Here, Algorithm~\ref{alg:0} suffers the sub-optimal choices commented in Section~\ref{sec:main} made to obtain a practical algorithm achieving non-trivial reconstruction when close to $\alpha_c(T,\eta)$. On the opposite, the static Bethe-Hessian of \cite{dall2019revisiting} is explicitly designed to optimally perform community detection for all values of $\alpha$ and any degree distribution, thereby justifying the two curves for large values of $\alpha$.

\begin{figure}[t!]
	\centering
	\includegraphics[width = \columnwidth]{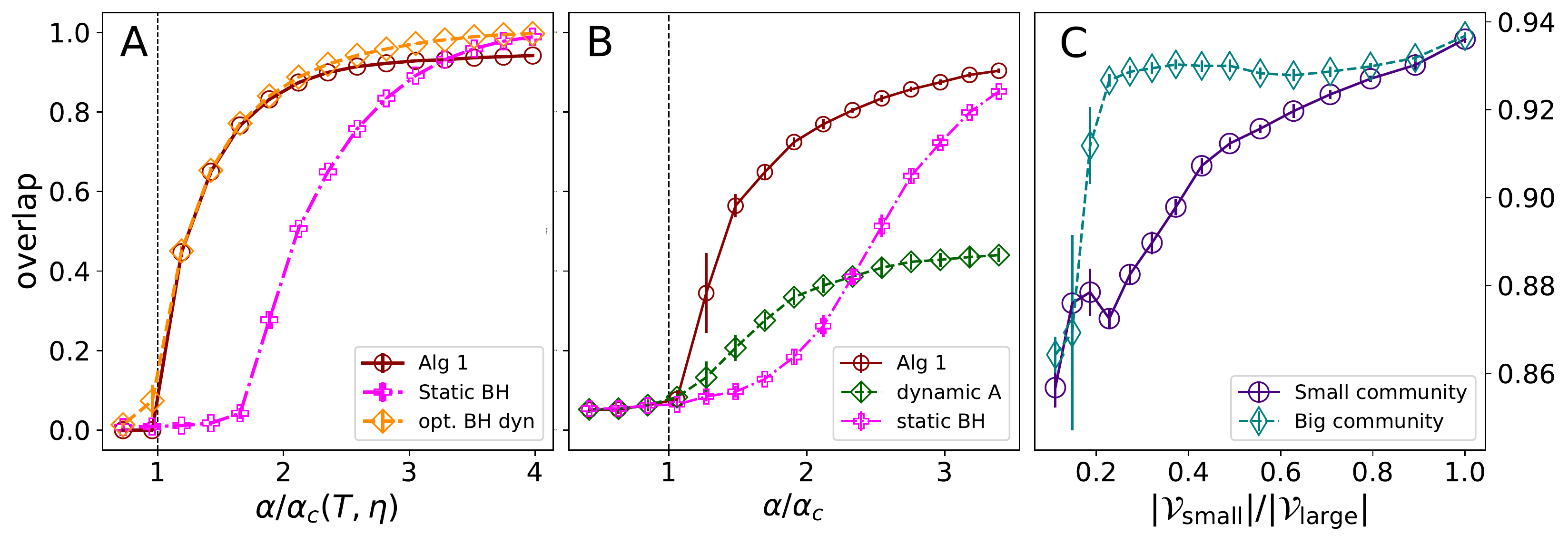}
	\caption{\textbf{A}: overlap comparison of Algorithm~\ref{alg:0} (Alg 1), the static Bethe-Hessian of \cite{dall2019revisiting} (Static BH) and the reconstruction obtained using the eigenvector with zero eigenvalue of $H_{\lambda,\eta}$ (opt. BH dyn). For this simulation  $n = 25~000$, $k = 2$, $\Phi = 1$, $c_{\rm out} = 0.5 \to 5$, $c = 6$, $T = 4$, $\eta = 0.9$. Averages are taken over $10$ samples. \textbf{B}: overlap comparison for Algorithm~\ref{alg:0} (Alg 1), the dynamical adjacency matrix of \cite{keriven2020sparse} (dyn A) and the static Bethe-Hessian of \cite{dall2019revisiting} (static BH) for large $T$. For this simulation $n = 500$, $k = 2$, $\Phi = 1$, $c_{\rm out} = 2 \to 5.5$, $c = 6$, $T = 25$, $\eta = 0.8$. \textbf{C}: Overlap averaged over time achieved by Algorithm~\ref{alg:0} on graphs with two communities of different size, as a function of the ratio of the size of the two communities. For this simulation $n = 10~000$, $T = 5$, $c = 6$, $\Phi = 1$, $\eta = 0.7$. The second largest eigenvalue of $C\Pi$ is fixed to $s_2(C\Pi) = 4$. Averages over $15$ samples.}
	\label{fig:optimal}
\end{figure}

More specifically, Figure~\ref{fig:optimal}.A confirms that one can devise an optimal (but impractical) algorithm that exploits the eigenvector of $H_{\lambda,\eta}$ with null eigenvalue, as suggested in Section~\ref{subapp:proof.prop1}. Close to the transition, the two dynamical methods perform similarly and largely outperform the static algorithm. For large values of $\alpha$, instead, Algorithm~\ref{alg:0} suffers the sub-optimal (but practical) choice of $\xi = \lambda_d$, while for $\xi = \lambda$ the dynamical Bethe-Hessian is never beaten by the static Bethe-Hessian.\\
Figure \ref{fig:optimal}.B instead compares the performance of Algorithm~\ref{alg:0} with the dynamical adjacency matrix \cite{keriven2020sparse} and the static Bethe-Hessian \cite{dall2019revisiting} for a large value of $T = 25$, well evidencing the advantage of finding a joint solution of the clustering problem at all times.

A last remark concerns the capability of Algorithm~\ref{alg:0} to recover communities of unequal sizes. Figure~\ref{fig:optimal}.C shows the accuracy of reconstruction of two communities of different size, as a function of the size of the smallest cluster over the size of the biggest.
In order to obtain comparable results for different values of the ratio of the sizes of the two clusters, the following strategy is adopted: let $\Pi \in \mathbb{R}^{2 \times 2}$ be the diagonal matrix defined so that $\Pi_{ii}$ is the fraction of nodes belonging to class $i$ (${\rm Tr}(\Pi) = 1$). By imposing $C\Pi \bm{1}_2 = c\bm{1}_n$, the expected average $c$ is independent of the class label and it corresponds to the leading eigenvalue of $C\Pi$. The second eigenvalue of $C\Pi$, instead, determines the hardness of the detection problem (in the case of two classes of equal size it equal $(c_{\rm in} - c_{\rm out})/2$). For a given ratio $\Pi_{11}/\Pi_{22}$, the matrix $C$ is constructed so to let the leading eigenvalue of $C\Pi$ equal to $c$, and the second eigenvalue equal to a fixed value. For each time $t\geq 2$, the size of each class is kept fixed, by reassigning the labels according to the rule
\begin{align}
\ell_{i_{t}} =   \begin{cases}
\ell_{i_{t-1}}  &\text{w.p.\@ } \eta \\
a &\text{w.p.\@ } (1-\eta)\Pi_{aa}, ~~a \in \{1,2\}.
\end{cases}
\end{align}

The overlap (averaged over time) is then evaluated independently over the large and small class, to keep this measure meaningful: in the case $|\mathcal{V}_{\rm small}|\gg |\mathcal{V}_{\rm large}|$, assigning all nodes to the same cluster would output a large overlap.

\section{A fast implementation}
\label{app:fast_algo}
A naive implementation of Algorithm~\ref{alg:0} runs in $\mathcal{O}(nT\sum_{l=k}^m l^2)$ where $m$ is the \emph{a priori} unknown number of negative eigenvalues of $H_{\lambda_d,\eta}$. Indeed, one (i) starts by computing the $k$ eigenvectors associated to the lowest eigenvalues of $H_{\lambda_d,\eta}$, costing $\mathcal{O}(nTk^2)$ via for instance classical restarted spectral Arnoldi algorithms~\cite{saad_numerical_2011}; (ii) verifies that the largest found eigenvalue is still negative; (iii) computes the $k+1$ eigenvectors associated to the lowest eigenvalues of $H_{\xi,\lambda_d}$; (iv) checks that the largest found eigenvalue is still negative; (v) iterates this process until the largest found eigenvalue crosses zero. 

A much faster approximate implementation is described in Algorithm~\ref{alg:fast}. The computation of the embedding $Y$ (line 10) should be done iteratively and thus costs $\mathcal{O}(pnT\log(nT))$, where $p$ indicates the order of the polynomial approximation $\tilde{f}$ (defined in line 8). The $T$ $k$-means steps cost $\mathcal{O}(nTk\log(nT))$. The overall cost is thus $\mathcal{O}(nTk\log(nT))$, where the constant $p$ is omitted as it is a problem-independent numerical factor.

To be complete, we recall here the two main arguments behind this accelerated algorithm: random projections and polynomial approximation. Further details  may be found in~\cite{tremblay_icassp16, ramasamy_compressive_2015, tremblay_icml16}.

\begin{algorithm}[t!]
	\begin{algorithmic}[1]
		\State \textbf{Input} : adjacency matrices $\{A^{(t)}\}_{t=1,\dots,T}$ of the undirected dynamical graph  $\mathcal{G}=\{\mathcal{G}_t\}_{t = 1,\dots,T}$, label persistence $\eta$, number of clusters $k$; and parameters typically set to $p=50$ (the order of the polynomial approximation) and $r=10\log(nT)$ (the dimension of the random projection)
		\For {$t = 1:T-1$}
		\State Remove from $A^{(t+1)}$ the edges appearing in both $A^{(t)}$ and $A^{(t+1)}$ (Appendix \ref{app:memory})
		\EndFor
		\State Compute $\lambda_d$ as in Algorithm~\ref{alg:0} and create the dynamical Bethe-Hessian matrix $H_{\lambda_d,\eta}\in\mathbb{R}^{nT\times nT}$
		\State Compute $\mu_{\rm min}$ and $\mu_{\rm max}$ the minimal and maximal eigenvalues of $H_{\lambda_d,\eta}$
		\State Build $H_{\lambda_d,\eta}' = H_{\lambda_d,\eta} -\mu_{\rm min} I$, the shifted positive semi-definite version of $H_{\lambda_d,\eta}$.
		\State Consider the step function $f(\mu)=1$ if $\mu\leq -\mu_{\rm min}$ and $0$ if $\mu>-\mu_{\rm min}$.
		\State Compute the coefficients $\{\alpha_k\}_{k=0,\ldots,p}$ of the order $p$ Jackson-Chebychev polynomial approximation of $f$ on the interval $[0, \mu_{\rm max}-\mu_{\rm min}]$: $$\forall \mu\in[0,\mu_{\rm max}-\mu_{\rm min}],\qquad f(\mu)\simeq \tilde{f}(\mu)=\sum_{k=0}^p \alpha_k \mu^k.$$
		\State Generate a random matrix $R\in\mathbb{R}^{nT \times r}$ with iid Gaussian entries such that $\mathbb{E}(RR^T)=I$. 
		\State Compute $Y\in\mathbb{R}^{nT\times r}$ as $$Y=\tilde{f}(H_{\lambda_d,\eta}')R = \sum_{k=0}^p \alpha_k H_{\lambda_d, \eta}'^k\; R$$
		\State Normalize the rows of $Y_{i,:} \leftarrow Y_{i,:}/{\Vert Y_{i,:} \Vert}$
		\For{$t =1:T$}
		\State Estimate the community labels $\{\hat{\ell}_{i_t}\}_{i = 1,\dots n}$ using ${k}$-class \emph{k-means} on the rows $\{Y_{i_t}\}_{i = 1,\dots,n}$.
		\EndFor
		\\
		\Return Estimated label vector $\hat{\bm{\ell}} \in \{1,\dots,{k}\}^{nT}$.
		\caption{A fast approximate implementation of Algorithm~\ref{alg:0}.}
		\label{alg:fast}
	\end{algorithmic}
\end{algorithm}

\noindent\textbf{A preliminary observation.} Let $X\in\mathbb{R}^{nT\times m}$ be the exact eigenvectors of $H_{\lambda_d,\eta}$ associated to negative eigenvalues. They are obviously also the eigenvalues between $0$ and $-\mu_{\rm min} > 0$ of the shifted matrix  (used in Algorithm~\ref{alg:fast}) $H'_{\lambda_d,\eta} = H_{\lambda_d,\eta} - \mu_{\rm min}I_{nT}$, where $\mu_{\rm min}$ is the smallest eigenvalue of $H_{\lambda_d,\eta}$. Algorithm~\ref{alg:0} then performs $k$-means on the rows of $\{X_{i_t}\}_{i = 1,\dots,n}$ for any $t=1,\ldots, T$. An important observation is that $k$-means only relies on the Euclidean distance between the feature vectors $\bm{f_i}=X^T\bm{\delta_{i_t}}\in\mathbb{R}^m$, where the only non-zero entry of $\bm{\delta_{i_t}} \in \{0,1\}^{nT}$ is precisely $i_t$,
\begin{align}
d_{ij}^2=||\bm{f_i}-\bm{f_j}||^2_2.
\end{align}
As such, $k$-means does not \emph{need} the exact matrix $X$, but rather only feature vectors whose interdistances verify the above. The random projections discussed in the next paragraph aim at creating random feature vectors whose interdistances concentrate around the above Euclidean distance.

\noindent\textbf{Random projection.} Denote by $R\in\mathbb{R}^{nT\times r}$ a random matrix with for example Gaussian i.i.d. entries verifying $\mathbb{E}(RR^T)=I$. 
Define $Y = XX^T R\in\mathbb{R}^{nT\times r}$ and new feature vectors $\bm{\bar{f}_{i}}=Y^T \bm{\delta_{i}}\in\mathbb{R}^{r}$. One has, denoting $\bm{\delta_{i j}}=\bm{\delta_{i}}-\bm{\delta_{j}}$:
\begin{align}
   \forall~ i,j\qquad \bar{d}^2_{ij} = ||\bm{\bar{f}_i}-\bm{\bar{f}_j}||^2_2 = ||R^T XX^T \bm{\delta_{ij}}||_2 
\end{align}
and in expectation:
\begin{align*}
    \forall~ i,j\qquad \mathbb{E}\left(\bar{d}^2_{ij}\right) &= \mathbb{E}\left(\bm{\delta_{ij}}^T XX^T RR^T XX^T     \bm{\delta_{ij}}\right)\\
    &=\bm{\delta_{ij}}^T XX^T \mathbb{E}\left(RR^T\right) XX^T \bm{\delta_{ij}}\\
    &=\bm{\delta_{ij}}^T XX^T XX^T \bm{\delta_{ij}}\\
    &=\bm{\delta_{ij}}^T XX^T \bm{\delta_{ij}}\\
    &=d^2_{ij}.
\end{align*}
Importantly, the concentration of the expectation around its expected value is \emph{fast}. The Jonhson Lindenstrauss lemma states that $r=\mathcal{O}(\frac{1}{\epsilon^2}\log{nT})$ suffices for a $(1+\epsilon)$ multiplicative approximation of the Euclidean distance (see~\cite{tremblay_icassp16,ramasamy_compressive_2015} for a lengthier discussion).

\noindent\textbf{Polynomial approximation.} In our context, these random projections are pointless as long as we do not have an efficient way to obtain $Y$ without actually computing $X$. This problem can be solved using a polynomial approximation. Let us write the diagonalized form of $H'_{\lambda_d,\eta}$ as $H'_{\lambda_d,\eta}=U\Lambda' U^T$ where $\Lambda'$ is the diagonal matrix of eigenvalues $\{\mu'_i\}$. Let us write the matrix function $f(H'_{\lambda_d,\eta})=Uf(\Lambda')U^T$ for any function $f$ defined on the spectrum of $H'_{\lambda_d,\eta}$. Let us consider the particular step-function function $f(\mu)$ that is equal to $1$ if $\mu\leq-\mu_{\rm min}$ and to $0$ if $\mu>-\mu_{\rm min}$. Note that $XX^T=f(H'_{\lambda_d,\eta})$.

Define $\tilde{f}(\mu)=\sum_{k=0}^p\alpha_k\mu^k$ a polynomial approximation of order $p$ of $f(\mu)$ on the interval $[0, \mu_{\rm max}-\mu_{\rm min}]$ (the larger $p$ the better the approximation). One can compute an approximation of $Y$ using $\tilde{f}$:
\begin{align*}
    Y &=  f(H'_{\lambda_d,\eta})\;R\\  
    &\simeq \tilde{f}(H'_{\lambda_d,\eta}) R = U\sum_{k=0}^p \alpha_k  \Lambda'^k U^T R = \sum_{k=0}^p \alpha_k H'^k_{\lambda_d,\eta}\;  R.
\end{align*}
The choice of which polynomial approximation to choose is not straightforward. One possible choice is to use Chebychev polynomials as they have a guarantee on the infinite norm of the approximation error. However, they tend to create Gibbs oscillation around sharp cut-offs of the function to approximate. As the function we wish to approximate here is a step function, it is customary to choose Jackson-Chebychev polynomials (which explicitly dampen these unwanted oscillations). See discussions in~\cite{tremblay_icml16,di2016efficient, JAY199921}. 

\begin{figure}
	\centering
	\includegraphics[width =\columnwidth]{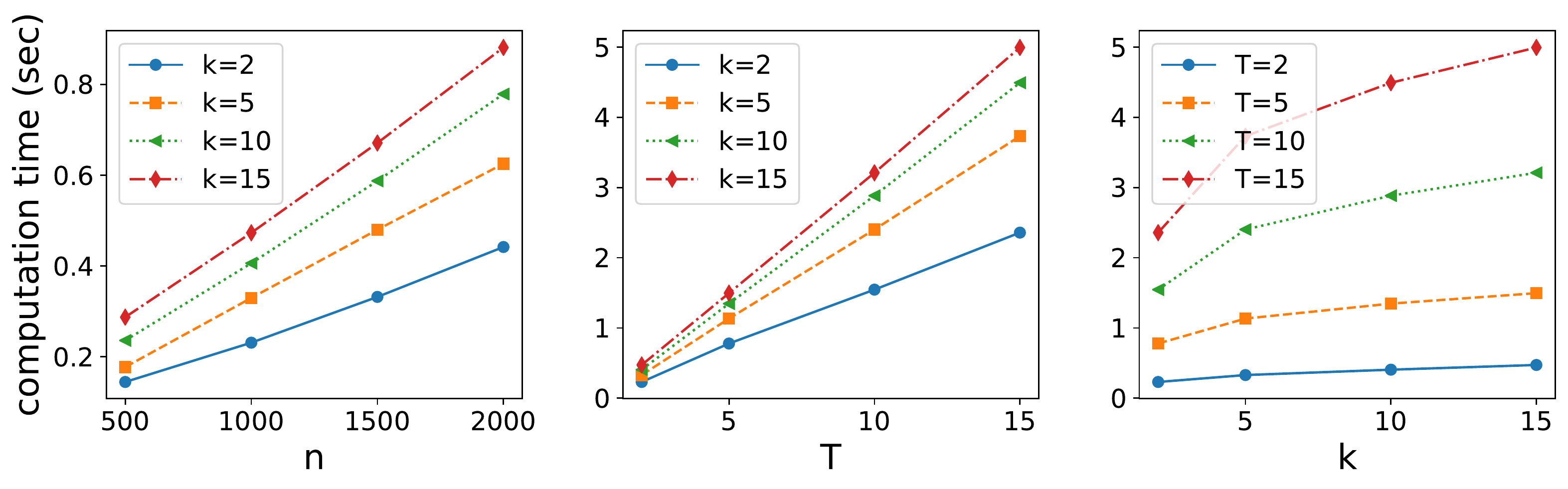}\\
	\includegraphics[width =\columnwidth]{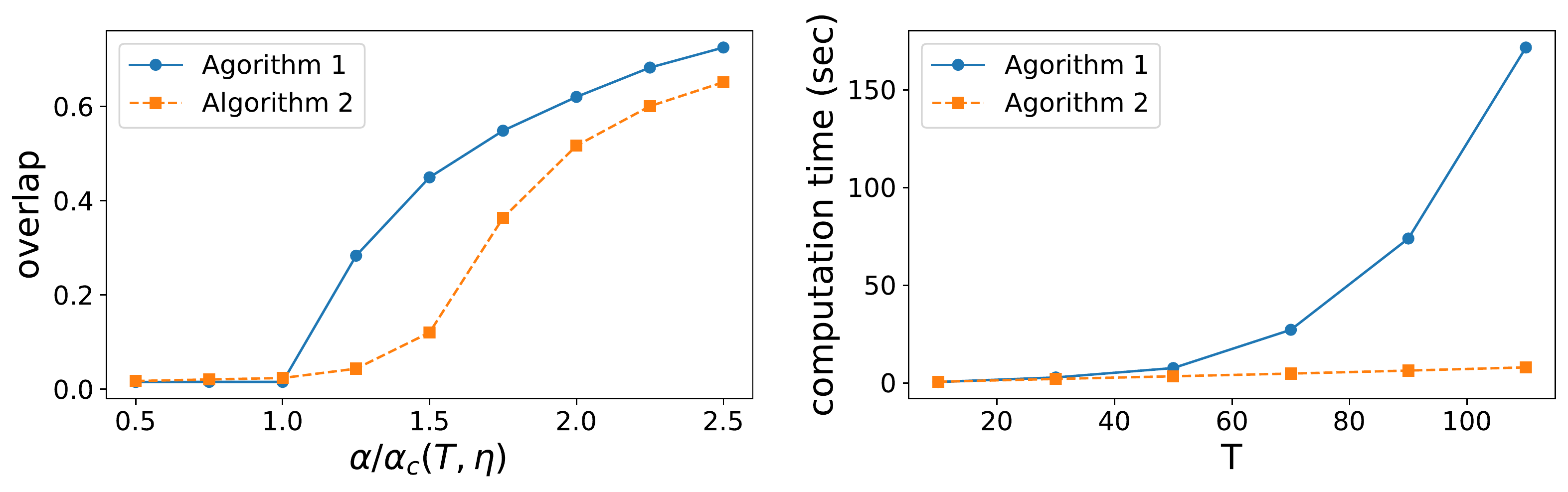}
	\caption{\textbf{Top}: computation time of Algorithm~\ref{alg:fast} versus the number of nodes $n$ (left, for $T=2$), the number of timesteps $T$ (middle, for $n=1000$) and the number of communities $k$ (right, for $n=1000)$, with parameters $\eta=0.7$, average degree $c=6$, $\Phi=1.6$, and $\alpha=1.5\alpha_c(T,\eta)$. \textbf{Bottom}: performance comparison between Algorithm~\ref{alg:0} and Algorithm~\ref{alg:fast} in terms of (left) overlap versus $\alpha/\alpha_c(T,\eta)$ for $n=5000$, $T=10$, $k=2$, $c=6$, $\eta=0.5$, $\Phi=1.6$ and in terms of (right) computation time versus $T$ for $n=300$, $k=2$, $c=6$, $\alpha=1.5\alpha_c(T,\eta)$, $\eta=0.5$, $\Phi=1.6$. On all figures, the results are the average over 40 experiments.}
	\label{fig:Algo2}
\end{figure}

\noindent\textbf{In practice.} Fig.~\ref{fig:Algo2} (top) experimentally illustrates that the complexity of Algorithm~\ref{alg:fast} is indeed linear in $n$, $T$ and $k$. The bottom of Fig.~\ref{fig:Algo2} compares both Algorithms in terms of overlap and computation time: Algorithm~\ref{alg:fast}, being only an approximation, never performs as well as Algorithm~\ref{alg:0}, especially as the detection problem becomes more difficult and the control parameter $\alpha$ approaches the transition point $\alpha_c$. However, the gain in computation time is drastic as $m$ increases (here $k$ is fixed to $2$ and $T$ increases). 

